\documentclass[aps,prb,showpacs,amsmath,amssymb,reprint,superscriptaddress]{revtex4-1}
\pdfoutput=1

\usepackage{graphicx}
\usepackage{color}
\usepackage{amsmath}
\usepackage{amssymb}
\usepackage{amsfonts}
\usepackage{hyperref}
\usepackage{bm}
\usepackage{xcolor}
\usepackage{soul}
\usepackage{cleveref}

\definecolor{PV-color}{rgb}{0.97,0.57,0.11}

\definecolor{FA-color}{rgb}{0.0,0.57,0.11}

\definecolor{TTH-color}{rgb}{0.0,0.0,1}

\begin{document}

\title{Domain wall motion in a diffusive weak ferromagnet
}

\author{Faluke Aikebaier}
\email[]{faluke.aikebaier@jyu.fi}
\affiliation{Department of Physics and Nanoscience Center, University of Jyv\"{a}skyl\"{a}, P.O. Box 35 (YFL), FI-40014 University of Jyv\"{a}skyl\"{a}, Finland
}

\author{Tero T. Heikkil\"{a}}
\email[]{tero.t.heikkila@jyu.fi}
\affiliation{Department of Physics and Nanoscience Center, University of Jyv\"{a}skyl\"{a}, P.O. Box 35 (YFL), FI-40014 University of Jyv\"{a}skyl\"{a}, Finland
}

\date{\today}

\begin{abstract}
We study the domain wall motion in a disordered weak ferromagnet, induced by injecting a spin current from a strong ferromagnet. Starting from the spin diffusion equation describing the spin accumulation in the weak ferromagnet, we calculate the force and torque acting on the domain wall. We also study the ensuing domain wall dynamics, and suggest a possible measurement method for detecting the domain wall motion via measuring the additional resistance.
\end{abstract}

\pacs{}
\keywords{}

\maketitle

\section{Introduction\label{sec:introduction}}

Current-driven domain wall motion has been an active field of research due to its applications in memory-storage devices~\cite{Parkin190}. Following a series of phenomenological theoretical works~\cite{doi:10.1063/1.324716,doi:10.1063/1.333530,doi:10.1063/1.351045,PhysRevB.54.9353} and experimental confirmations,~\cite{doi:10.1063/1.1507820,doi:10.1063/1.1594841,doi:10.1063/1.1588736,Vernier_2004,PhysRevLett.92.077205} a microscopic theory of domain wall motion was presented more than a decade ago.~\cite{PhysRevLett.92.086601} The essential mechanism of such effects is the transfer of momentum and spin to the local magnetization due to a force and a (spin) torque, respectively, exerted by a spin polarized current passing through the domain wall.~\cite{TATARA2008213} However, spin-polarized currents may reduce the spin torque efficiency with an increasing temperature due to Joule heating.~\cite{PhysRevLett.97.046602,doi:10.1063/1.2709989} 

One suggestion to reduce the Joule heating is to replace the spin-polarized charge current with the pure spin current to induce the domain wall motion. Such pure spin currents have been realized in a lateral spin valve geometry,~\cite{PhysRevLett.105.076601,PhysRevB.88.214405,Pfeiffer_2017} see for example Fig.~\ref{fig:schematics}. The scenario in this case is as follows. A spin polarized current is injected from a ferromagnet to a nonmagnetic material, transported and absorbed by the second ferromagnet containing a domain wall. The absorbed pure spin current then induces a domain wall motion. It was shown that the domain wall motion in this case is also very efficient, in terms of the change of the magnetization at the interface of the ferromagnet where the spin current is absorbed. The force and torque in this structure have also been calculated for a case of weak impurity scattering,~\cite{PhysRevB.87.094404} but the ensuing domain wall dynamics have not yet been studied theoretically.  

One important feature of the pure spin current compared to the spin-polarized current is that it decays within a length scale called spin-relaxation length, due to the spin-relaxation processes. In fact, spin relaxation significantly affects the current-driven domain wall motion.~\cite{TATARA2008213} For example, the spin relaxation of conduction electrons is one of the most relevant mechanisms for the damping of the domain wall motion. Moreover, it enhances the nonadiabaticity parameter of the domain walls close to the adiabatic limit.~\cite{PhysRevLett.93.127204,Thiaville_2005} In disordered ferromagnets, it has also been shown that the domain wall motion is very efficient even in the case of weak ferromagnetism with low spin polarization.~\cite{PhysRevB.80.184406} Therefore, studying the domain wall dynamics in the presence of pure spin current without the accompanied charge current may give rise to interesting new features. 

Here we consider a similar structure with the one in Ref.~\onlinecite{PhysRevB.87.094404}, except that the nonmagnetic metal is replaced by a weak ferromagnet containing a domain wall,  and a spin polarized current is injected from a strong ferromagnetic electrode. We define the concepts of the "weak" and "strong" ferromagnets based on the size of the spin polarization and the possibility of using the spin diffusion equation to describe the two systems. In particular, in the strong ferromagnet we assume a spin-polarized Fermi surface, described by spin-dependent densities of states $N_\sigma$, diffusion constants $D_\sigma$ and conductivities $\sigma_\sigma = e^2 N_\sigma D_\sigma$.\cite{PhysRevB.48.7099} In this case, we can study the spin polarized current in a homogeneous ferromagnet by writing diffusion equations separately for the two spin bands. On the other hand, the weak ferromagnet has a weakly spin-split Fermi surface (small exchange field) for which $\sigma_\uparrow=\sigma_\downarrow$. In this case we can include the Hanle precession term into the kinetic equations and therefore rigorously describe spin accumulation in the case of an inhomogeneous magnetization.

The spin polarized current injected from the strong ferromagnetic electrode creates a spin accumulation in the weak ferromagnet which decays exponentially due to the spin relaxation processes. This spin accumulation can be described by a spin diffusion equation with spin independent parameters, and it describes a spin current in a disordered wire. The solutions for the position dependent spin accumulation around the domain wall allows us to compute the force $f$ and torque $\tau_z$ on the domain wall residing at a distance $X$ from the injector. We show that they are characterized by three length scales: domain wall size $\lambda$, spin relaxation length $\ell_{s}$, and the magnetic length $\ell_h$. These length scales can in principle show up in any order, and we find how the force and torque depend on the order of those scales. In particular, due to the spin relaxation both the force and torque are exponentially decaying as functions of the distance of the domain wall from the injector, similar to the case in Ref.~\onlinecite{PhysRevB.87.094404}. We also study the resulting domain wall dynamics, and show that the domain wall motion with decaying force and torque has its characteristic features. In particular, the dynamics can cross between different dynamic regimes depending on the position of the domain wall, and depending on the hierarchy of the length scales affecting the relative size of force and torque: In the case of a large torque and weak force, the domain wall motion can cross over from the unpinned motion for $\tau_z(X) > k_\perp \alpha_0$ to the limit of intrinsic pinning with $\tau_z(X) < k_\perp \lambda$, eventually stopping the domain wall. Here $k_\perp$ is a quantity characterizing the hard-axis anisotropy. On the other hand, if the force dominates and is large enough close to the injector, there is a crossover between oscillatory dynamics for $f(X) > \alpha_0 k_\perp$ and linearly (in time) decaying dynamics for $f(X) < \alpha_0 k_\perp$. Here $\alpha_0$ describes damping.

We also suggest a possible measurement of the domain wall motion via the changes in the injection resistance, linked to the dependence of the injection resistance on the local spin accumulation at the position of the contact. Since the latter depends on the position of the domain wall, so does the injection resistance.

The outline of the paper is as follows. We first introduce the model, a weak ferromagnet containing a domain wall in contact with a spin-polarized ferromagnetic injector, in Sec.~\ref{sec:ModelandMethod}. We also solve the spin diffusion equation with proper boundary conditions which describes the spin accumulation in this model. The force and torque due to the spin current are calculated in Sec.~\ref{sec:Forceandtorque}. We study the domain wall dynamics in Sec.~\ref{sec:Domainwalldynamics}, and the possible measurement method accessing this dynamics in Sec.~\ref{sec:InjectionResistance} before the conclusions in Sec.~\ref{sec:conclusion}.

\section{Model and Method\label{sec:ModelandMethod}}

\begin{figure}[t]
\centering
\includegraphics[width=1\linewidth]{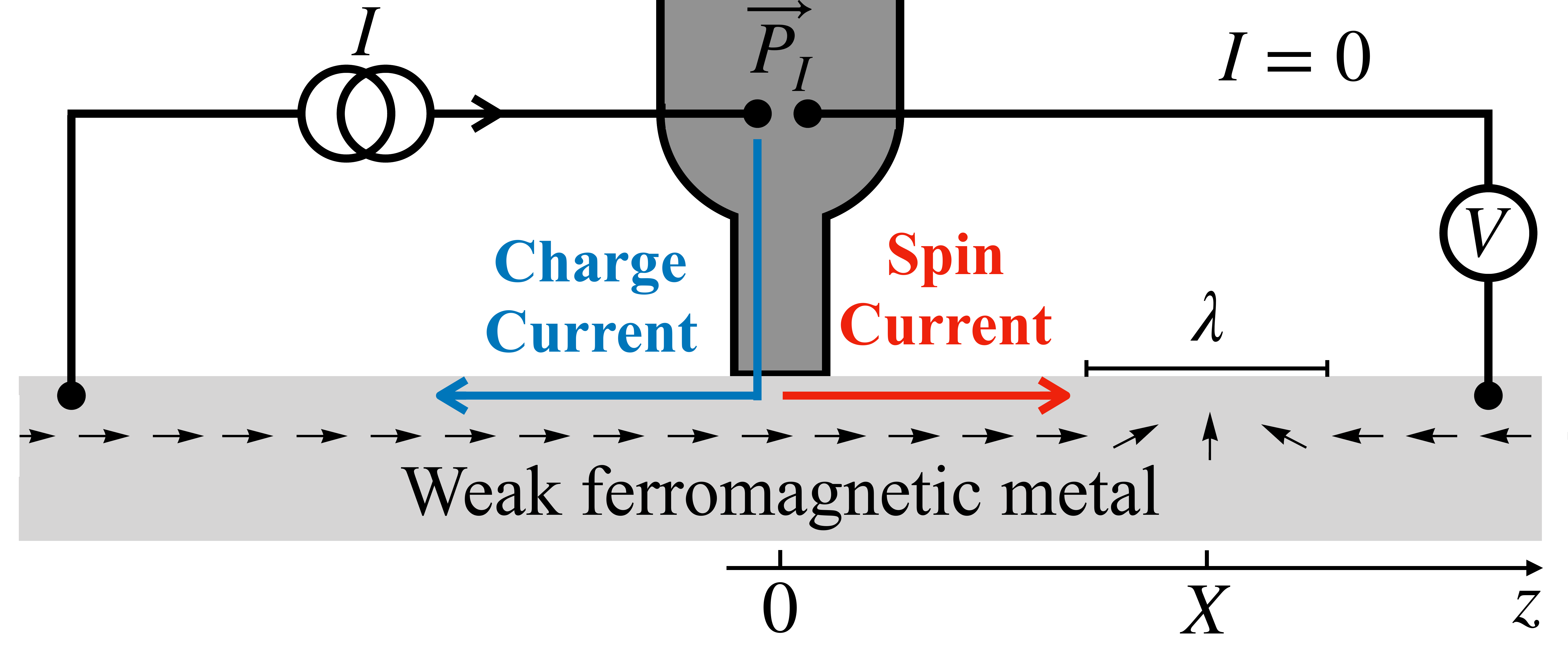}
\caption{\label{fig:schematics} Schematic view of the structure considered in this paper. A spin polarized current is injected from a strong ferromagnet to a diffusive weak ferromagnet containing a domain wall.}
\end{figure}

We study the domain wall motion in the structure in Fig.~\ref{fig:schematics}. A spin polarized current is injected from a strong ferromagnet to a diffusive weak ferromagnet containing a domain wall. The injected current circulates on the left side of the injector, and a spin accumulation is induced in the weak ferromagnet. The decaying spin accumulation results in a spin current in both directions, capable of inducing a force and a torque on the domain wall.

On the right side of the injector, the weak ferromagnet contains a domain wall, and the magnetization is inhomogeneous. The inhomogeneity is shown in the exchange field as
\begin{equation}\label{eq:ExchangeField}
\boldsymbol{h}=h(\sin\theta\cos\phi,\sin\theta\sin\phi,\cos\theta),
\end{equation}
where $h$ is the strength of the exchange splitting. Here $\theta$ and $\phi$ are the in-plane and out-of-plane components of the magnetization angle. For domain wall motion, $\phi$ is only a function of time,~\cite{PhysRevLett.92.086601} and the rotation is described by the angle $\theta$. A N\'eel domain wall is energetically favoured in thin films, namely, the rotation of the magnetization happens in the plane of the domain wall ($\phi=0$). Then $\theta$ can be expressed by a variational ansatz~\cite{PhysRevB.99.104504}
\begin{align}\label{eq:RotationAngle}
\begin{split}
&\theta(z)=\pi\Theta\left(z-X-\frac{\lambda}{2}\right)\\
&+\frac{\pi}{\lambda}\left(z-X+\frac{\lambda}{2} \right)\Theta\left(z-X+\frac{\lambda}{2} \right)\Theta\left(X+\frac{\lambda}{2}-z \right) ,
\end{split}
\end{align}
where $\Theta(z)$ is the Heaviside step function, $X$ is the position of the domain wall center, and $\lambda$ is the domain wall size. The variational ansatz to the rotation angle, instead of the typically used hyperbolic functions~\cite{TATARA2008213} with slightly lower domain wall energy, brings certain conveniences to the analytical treatment of the problem while capturing the essential physics of the domain wall. Since the derivative of $\theta(x)$ is a constant inside the domain wall, the spin diffusion equation, which describes the nonequilibrium spin accumulation, can be simplified [see Eq.~\eqref{eq:SpinDiffEqnUnrotated}]. The nonanalyticity of the derivative of $\theta(x)$ at the domain wall boundary can be transformed into boundary conditions of the spin diffusion equation [see Eq.~\eqref{eq:BoundaryConditionsContinuity1} to Eq.~\eqref{eq:BoundaryConditionsContinuity3}].

The spin accumulation in the weak ferromagnet is described by a spin diffusion equation in Eq.~\eqref{eq:SpinDiffusionEquation}. With the domain wall structure in Eq.~\eqref{eq:RotationAngle}, it can be written as
$$
\hbar D\partial_z^2 \boldsymbol{s}=\frac{\hbar}{\tau_{s}}\boldsymbol{s}-2\boldsymbol{h}\times\boldsymbol{s},
$$
where $D$ is the diffusion constant, $\tau_s$ is the spin-flip relaxation time, and  $\boldsymbol{s}=(s_1,s_2,s_3)$ is a spin accumulation vector. The spin-relaxation length is defined as $\ell_s=\sqrt{D\tau_s}$.

We can use an SU(2) gauge transformation to treat the exchange field as homogeneous. We define a rotation matrix as
$$
\hat{R}=e^{i\sigma_2\theta/2}e^{i\sigma_3\phi/2},
$$
so that we can write the spin accumulation as
\begin{equation}\label{eq:UnrotSpinAccumulation}
\boldsymbol{s}\cdot\boldsymbol{\sigma}=\hat{R}^\dagger\boldsymbol{s}_0\cdot\boldsymbol{\sigma}\hat{R},
\end{equation}
where $\boldsymbol{\sigma}=(\sigma_1,\sigma_2,\sigma_3)$ is a vector of Pauli spin matrices. Here the rotated spin accumulation $\boldsymbol{s}_0=(s_1^0,s_2^0,s_3^0)$ satisfies the following spin diffusion equation
\begin{equation}\label{eq:SpinDiffEqnUnrotated}
\hbar D\hat{\partial}_z^2 \boldsymbol{s}_0=\frac{\hbar}{\tau_{s}}\boldsymbol{s}_0-2h\hat{z}\times\boldsymbol{s}_0,
\end{equation}
where $\hat{z}=(0,0,1)$, $\hat{\partial}_z\boldsymbol{Y}=\partial_z\boldsymbol{Y}+\partial_z\theta(z)(\hat{y}\times\boldsymbol{Y})$ from the fact that $\hat{\partial}_z\boldsymbol{Y}\cdot\boldsymbol{\sigma}=\partial_z\boldsymbol{Y}\cdot\boldsymbol{\sigma}-[A, \boldsymbol{Y}\cdot\boldsymbol{\sigma}]=\partial_z\boldsymbol{Y}\cdot\boldsymbol{\sigma}+\partial_z\theta(x)(\hat{y}\times\boldsymbol{Y})\cdot\boldsymbol{\sigma}$, in which $A=i\sigma_2\partial_z\theta(z)/2$ is an SU(2) type vector potential, $\boldsymbol{Y}=\boldsymbol{s}_0$ or $\partial_x\boldsymbol{s}_0$, and $\hat{y}=(0,1,0)$. The derivative of $\theta(z)$ divides the weak ferromagnet into three regions. In the domain wall region it is a constant, and to the left and the right sides of the domain wall region, $\theta'(z)=0$. However, $\theta'(z)$ is discontinuous at the boundary of the domain wall. Therefore, we need a boundary condition to describe a continuous spin accumulation.

We can integrate Eq.~\eqref{eq:SpinDiffEqnUnrotated} at the boundary of the domain wall, and obtain the boundary conditions 
\begin{equation}\label{eq:BoundaryConditionsContinuity1}
\partial_z s_1^0\vert_{z_b^{\pm}}-\partial_z s_1^0\vert_{z_b^{\mp}}=-\frac{\pi}{\lambda}s_3^0\vert_{z_b^{\pm}},
\end{equation}
\begin{equation}\label{eq:BoundaryConditionsContinuity2}
\partial_z s_2^0\vert_{z_b^{\pm}}-\partial_z s_2^0\vert_{z_b^{\mp}}=0,
\end{equation}
\begin{equation}\label{eq:BoundaryConditionsContinuity3}
\partial_z s_3^0\vert_{z_b^{\pm}}-\partial_z s_3^0\vert_{z_b^{\mp}}=\frac{\pi}{\lambda}s_1^0\vert_{z_b^{\pm}}.
\end{equation}
At the domain wall edges $z=z_b^{\pm}=\pm(X\pm\lambda/2)$, and $\pm$ refers to the right and left sides of the domain wall boundary. 

The second group of boundary conditions represent the injection of the spin polarized current. As we show in Appendix \ref{sec:Current throughtheinjector}, the spin injection from a contact with a strong ferromagnet with magnetization oriented in the $z$ direction and biased with potential $V$ can be described with the spin currents at the injection point,
\begin{equation}\label{eq:Boundarycondition1}
\hbar D\partial_z s_1^0=0
\end{equation}
\begin{equation}\label{eq:Boundarycondition2}
\hbar D\partial_z s_2^0=0
\end{equation}
\begin{equation}\label{eq:Boundarycondition3}
\hbar D\partial_z s_3^0=k_I\hbar D(s_3^0-P_I\gamma VN_0),
\end{equation}
where $k_I$ is an injector transparency, $P_I$ is an injector polarization (see Appendix \ref{sec:Current throughtheinjector} for precise definitions of these quantities in terms of the properties of a ferromagnetic injector wire), $V$ is the voltage at the injector, and $N_0$ is the density of states at the Fermi level. The voltage is rescaled by a factor $\gamma$ [defined in Eq.~\eqref{eq:RescalingFactorVoltage}], due to fact that the spin accumulation in the weak ferromagnet is affected by the spin accumulation in the injector, see the details in Appendix \ref{sec:Current throughtheinjector}.

Making the equations dimensionless, we find that the domain wall physics is here described by three length scales: (i) domain wall size $\lambda$, (ii) spin relaxation length $\ell_s$, and (iii) the magnetic length $l_h=\sqrt{\hbar D/h}$. The latter indicates the length within which a non-collinear component of the spin accumulation rotates a full period around the local magnetization direction. This is an important scale since both the force and the torque depend on such non-collinear components, as shown in Eqs.~(\ref{eq:Force_spinaccumulation},\ref{eq:Torque_spinaccumulation}).

The "phase diagram" of different dynamical regimes depends on two dimensionless parameters corresponding to the ratios of these scales. In addition, the injector spin polarization $P_I$ describes the efficiency of spin injection (the size of spin current for a given amount of charge current), whereas the interface transparency parameter $k_I$ determines how strongly the resistance of the injector depends on the domain wall position.

In many strong ferromagnetic metals like iron and cobalt, the exchange splitting $h$ is of the order of $1\ \rm{eV}$.~\cite{WOHLFARTH19801} This then leads to a very small $l_h$, of the order of the atomic lattice spacing. For a weak ferromagnet, for example CuNi, it is around $0.05\ \rm{eV}$.~\cite{oboznov2006thickness} This leads to a magnetic length $l_h$ between $10$ to $25\ \rm{nm}$.~\cite{oboznov2006thickness,PhysRevLett.86.2427} On the other hand, depending on the exact materials or sample properties (e.g. thickness and concentration of Ni), the domain wall size $\lambda$ and the spin-relaxation lengths $\ell_s$ of CuNi range from $15$ to $25\ \rm{nm}$~\cite{doi:10.1063/1.4979267} (estimated from measured anisotropy energy and exchange stiffness constants) and from $7$ to $25\ \rm{nm}$~\cite{PhysRevB.54.9027}, respectively. This yields $\lambda/l_h\sim0.5...1.5$ and $l_h/\ell_s\sim0.4...3.6$. As there are also other materials with weak ferromagnetism, we also cannot exclude the other possibilities. In order to understand various properties of the domain wall motion induced from a spin current, we also consider these ratios outside of these ranges in the following discussions.

\begin{figure}[t]
\centering
\includegraphics[width=1\linewidth]{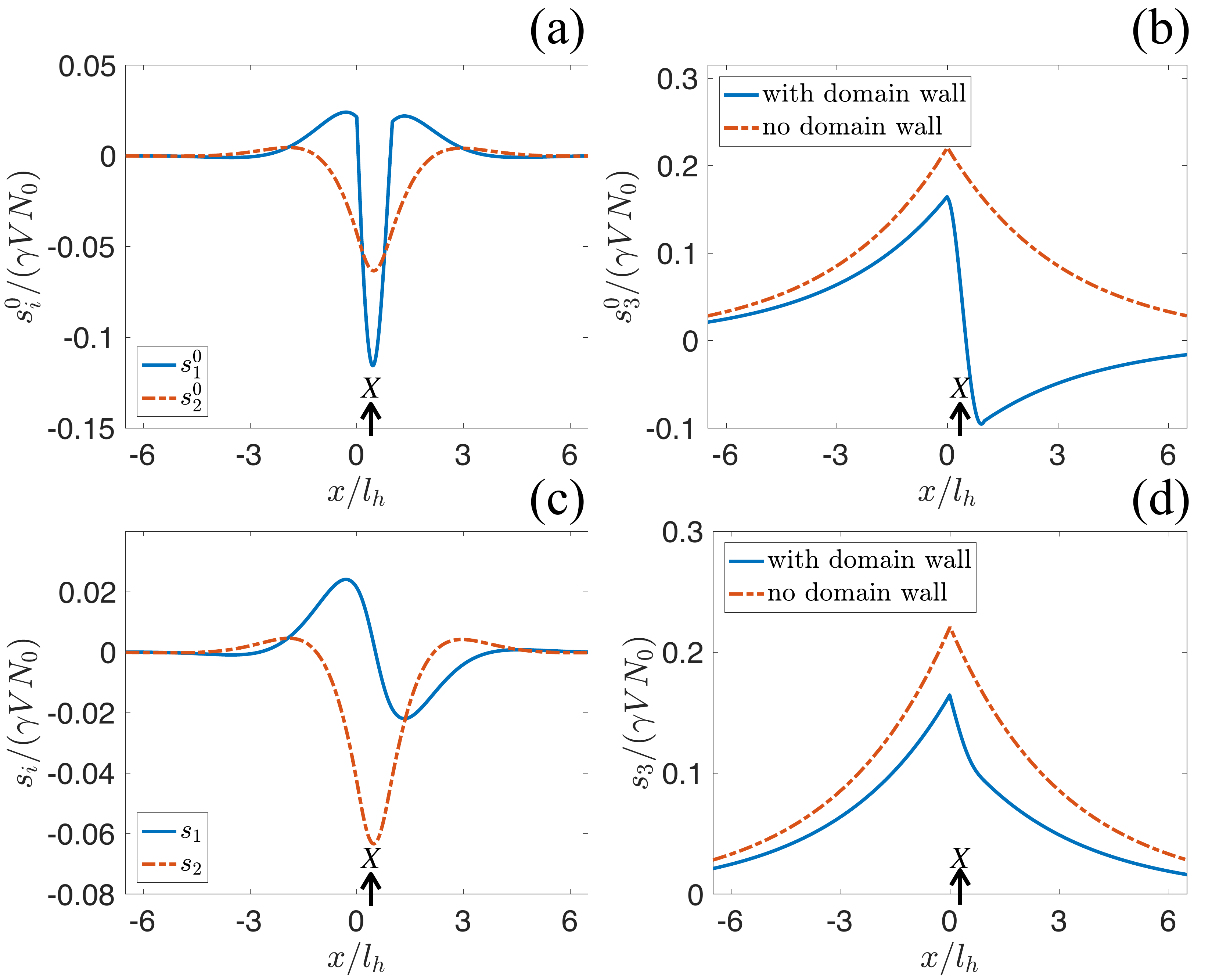}
\caption{\label{fig:solspindiffusioneqn} Solutions of the spin diffusion equation. The solutions in the rotated space are shown in (a) and (b), and in the unrotated space are shown in (c) and (d). We also compare $s_3^0$ and $s_3$ with the spin accumulation in the case of homogeneous magnetization (no domain wall). Here the results are plotted for $\ell_s=3.2l_h$, $P_I=0.5$, $k_Il_h=0.5$, and $\lambda=l_h$.
The injector is placed at $x=0$, whereas the domain wall center is at $X=0.5\lambda$ marked in the figure.}
\end{figure}

With the boundary conditions in  Eq.~\eqref{eq:BoundaryConditionsContinuity1} to Eq.~\eqref{eq:BoundaryConditionsContinuity3} and in Eq.~\eqref{eq:Boundarycondition1} to Eq.~\eqref{eq:Boundarycondition3}, we can solve the rotated spin diffusion equation in Eq.~\eqref{eq:SpinDiffEqnUnrotated}. They can be solved analytically (see  Appendix~\ref{sec:SpinAccuInInhomoMagnt}), but the solutions are in general quite lengthy. Rather, we plot the components of the spin accumulation for an example set of parameters as a function of position in Fig.~\ref{fig:solspindiffusioneqn}(a,b). We can see that $s_1^0$ is a monotonously increasing (decreasing) function of position in region to the left (right) side of the domain wall, and reaches a minimum in the domain wall center. The second component of spin accumulation $s_2^0$ smoothly goes to zero away from the domain wall center. Compared to the spin accumulation in the case without the domain wall, $s_3^0$ changes sign in the domain wall region and exponentially decreases in region to the right of the domain wall. 

The unrotated spin accumulation is given by Eq.~\eqref{eq:UnrotSpinAccumulation}. More specifically, we can write
\begin{equation}\label{eq:UnrotatedSpinAccumulation1}
s_1=\cos\phi(s_1^0\cos\theta+s_3^0\sin\theta)-s_2^0\sin\phi
\end{equation}
\begin{equation}\label{eq:UnrotatedSpinAccumulation2}
s_2=s_2^0\cos\phi+\sin\phi(s_1^0\cos\theta+s_3^0\sin\theta)
\end{equation}
\begin{equation}\label{eq:UnrotatedSpinAccumulation3}
s_3=s_3^0\cos\theta-s_1^0\sin\theta.
\end{equation}
The unrotated components of the spin accumulation are plotted for $\phi=0$ in Fig.~\ref{fig:solspindiffusioneqn}(c,d). Compared to the rotated solution, $s_2$ remains the same but $s_1$ changes sign on the two sides of the domain wall center, and $s_3$ also makes a difference compared to the case without the domain wall. In the next section, we use these spin accumulations to calculate the force and torque.

\section{Force and torque\label{sec:Forceandtorque}}
The force and torque acting on the domain wall are given by\cite{PhysRevLett.92.086601,TATARA2008213}
\begin{equation}\label{eq:Force}
F=-\int d^3z\nabla\boldsymbol{h}\cdot\boldsymbol{s}
\end{equation}
\begin{equation}\label{eq:Torque}
T_z=-\int d^3z(\boldsymbol{h}\times\boldsymbol{s})_z,
\end{equation}
where exchange field $\boldsymbol{h}$ is given in Eq.~\eqref{eq:ExchangeField}, and the components of the spin accumulation $\boldsymbol{s}=(s_1,s_2,s_3)$ are shown in Eq.~\eqref{eq:UnrotatedSpinAccumulation1} to Eq.~\eqref{eq:UnrotatedSpinAccumulation3}. Substituting these to the force and torque in Eq.~\eqref{eq:Force} and Eq.~\eqref{eq:Torque}, we obtain
\begin{equation}\label{eq:Force_spinaccumulation}
F=-\frac{h\pi W}{\lambda}\int dz s_1^0
\end{equation}
\begin{equation}\label{eq:Torque_spinaccumulation}
T_z=-h W\int dz s_2^0\sin\theta,
\end{equation}
where $W$ is the cross sectional area of the weak ferromagnet. 

\begin{figure}[h]
\centering
\includegraphics[width=1\linewidth]{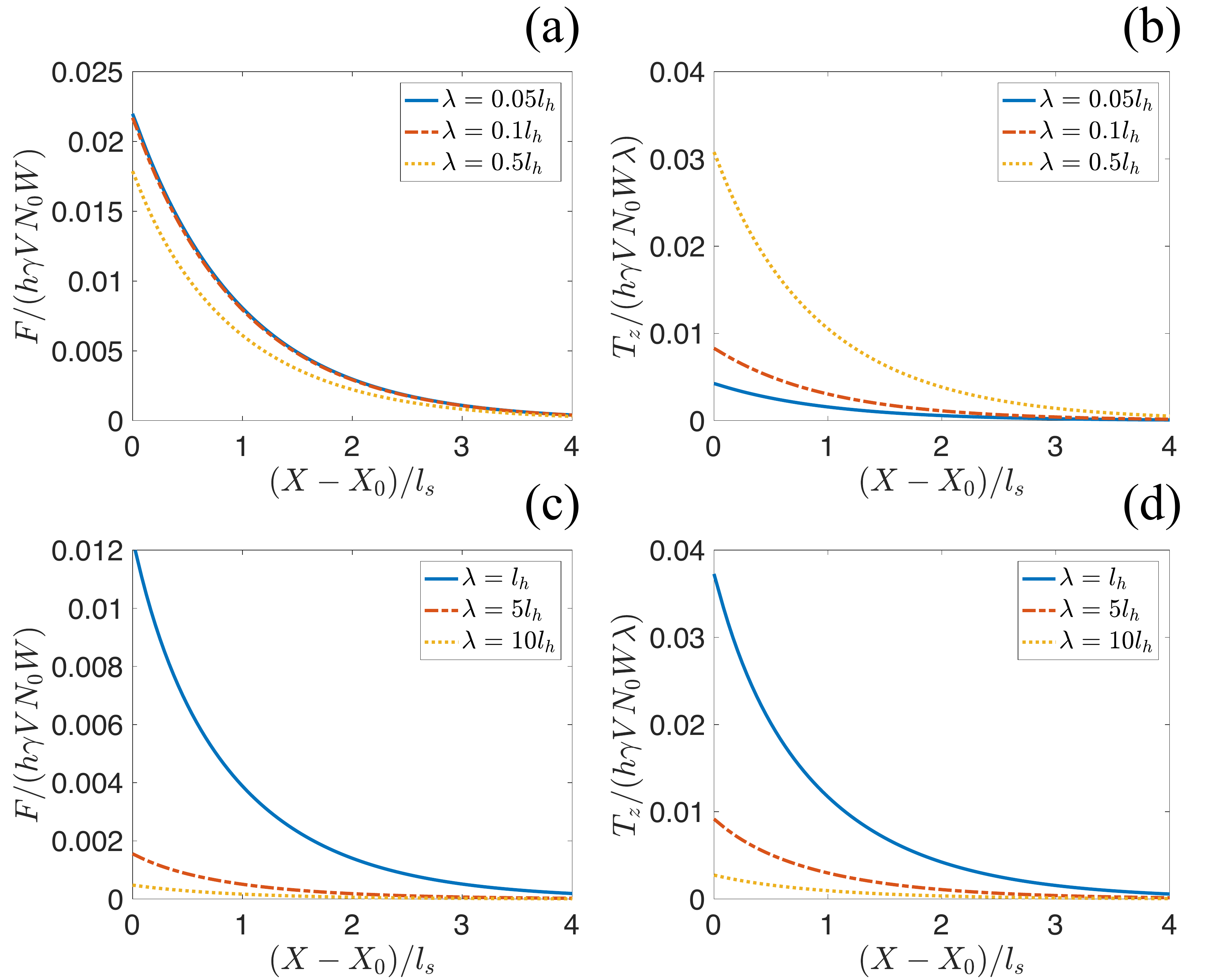}
\caption{\label{fig:force_torque} Force and torque for smaller domain walls in (a,b), and for larger domain walls in (c,d), as a function of domain wall center position $X-X_0$, where $X_0=\lambda/2$ is the shortest distance of the domain wall center to the right of the injector. 
Here the results are plotted for $\ell_s=3.2l_h$, $P_I=0.5$, and $k_Il_h=0.5$.}
\end{figure}

The force and torque as a function of the domain wall position $X$ are plotted in Fig.~\ref{fig:force_torque} for a few sets of parameters. 
The common feature of all the cases are that both decay exponentially as a function of $X$. This is due to the fact that the spin accumulation and the resulting spin current, which induces the domain wall motion, decays exponentially within the spin-relaxation length $\ell_s$. These features are also very similar to the ones in Ref.~\onlinecite{PhysRevB.87.094404}.  From Fig.~\ref{fig:force_torque}(a,c), we can see that the force is independent of the domain wall size for small domain walls, and it is smaller for larger domain walls. On the other hand, the torque has a nonmonotonic dependence on the domain wall size $\lambda$, as shown in Fig.~\ref{fig:force_torque}(b,d). It first increases as $\lambda$ increases up to of the order of $l_h$, and then becomes smaller for larger domain walls. This is not the same with the case of current driven domain wall motion, where the torque is much larger than the force for larger domain walls.~\cite{TATARA2008213} This is due to the fact that when a spin relaxation length $\ell_s$ is smaller than the domain wall size $\lambda$ ($\ell_s<\lambda$), due to the decaying spin current, less spins are transferred to the domain wall. This results in the smaller torque for larger domain wall sizes in Fig.~\ref{fig:force_torque}(d).

\begin{figure}[t]
\centering
\includegraphics[width=1\linewidth]{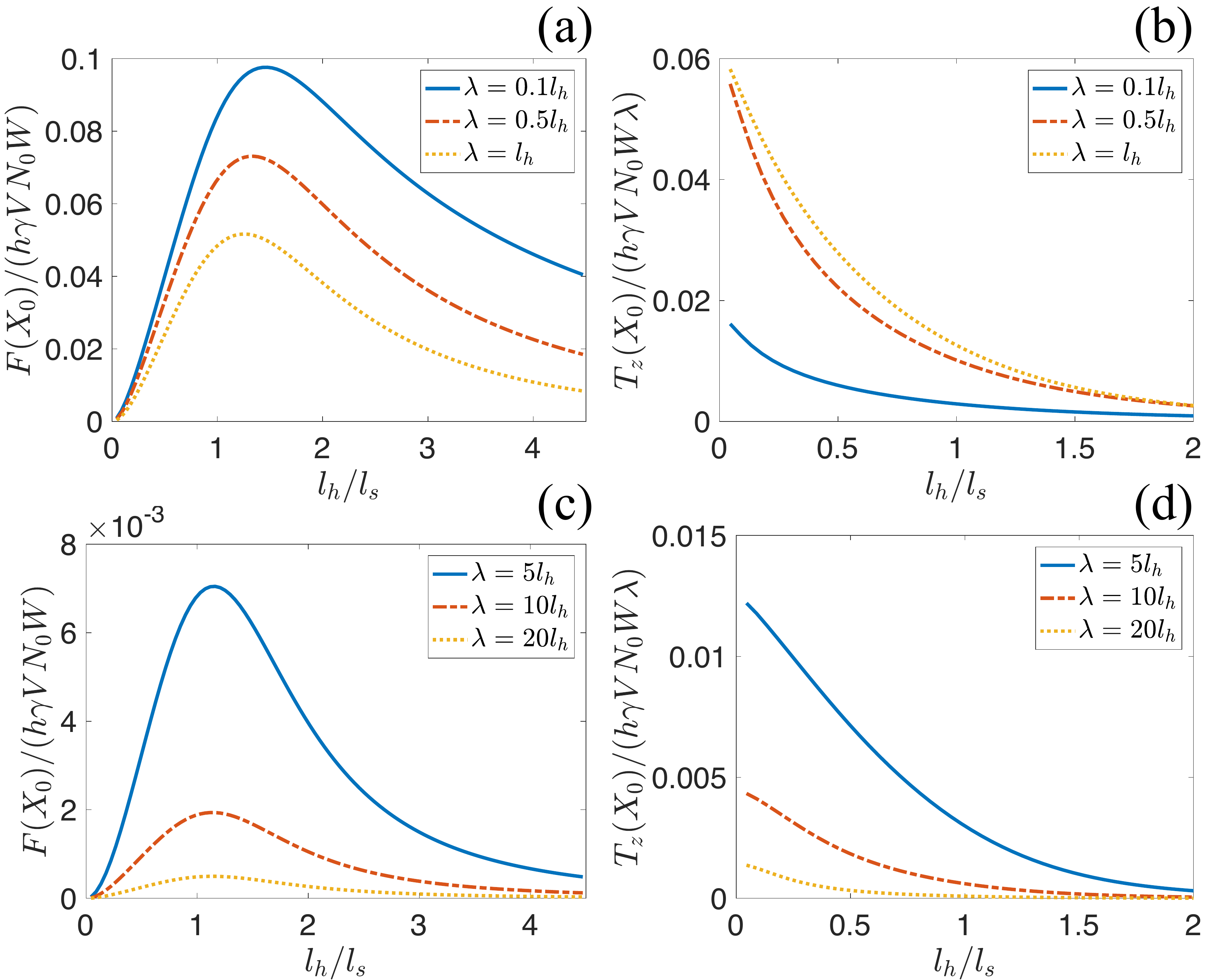}
\caption{\label{fig:force_torque_2} Force and torque for smaller domain walls in (a,b), and for larger domain walls in (c,d), as a function of inverse spin relaxation length $\ell_s$. Here the force and the torque are plotted for the domain wall position $X_0=\lambda/2$. The parameters used in the calculations are $P_I=0.5$ and $k_Il_h=0.5$.}
\end{figure}

The dependence of the force and the torque on the spin relaxation length are shown in Fig.~\ref{fig:force_torque_2}. We can see that the torque is a monotonously decreasing function of the inverse relaxation length, i.e., decreasing spin relaxation increases the torque, as expected from the fact that torque results from spin transfer. On the other hand, the force is a non-monotonic function of $l_h/\ell_s$. It also decays if the spin relaxation becomes strong (i.e., $l_h \gg \ell_s$). However, it also becomes small for a small magnetic length 
$l_h \ll l_s$. This is due to the fact that contrary to the torque, which within our model only comes from the domain wall region (that is where $\theta \neq 0$ in Eq.~\eqref{eq:Torque_spinaccumulation}), the force depends on the spin accumulation component $s_1^0$ also around the domain wall. However, for small $l_h$, this component oscillates rapidly, and thus the average force becomes small. Analogously, both the force and the torque become smaller for larger $\lambda/l_h$. This is due to the oscillations of the spin accumulation inside the domain wall region.

\begin{figure}[h]
\centering
\includegraphics[width=1\linewidth]{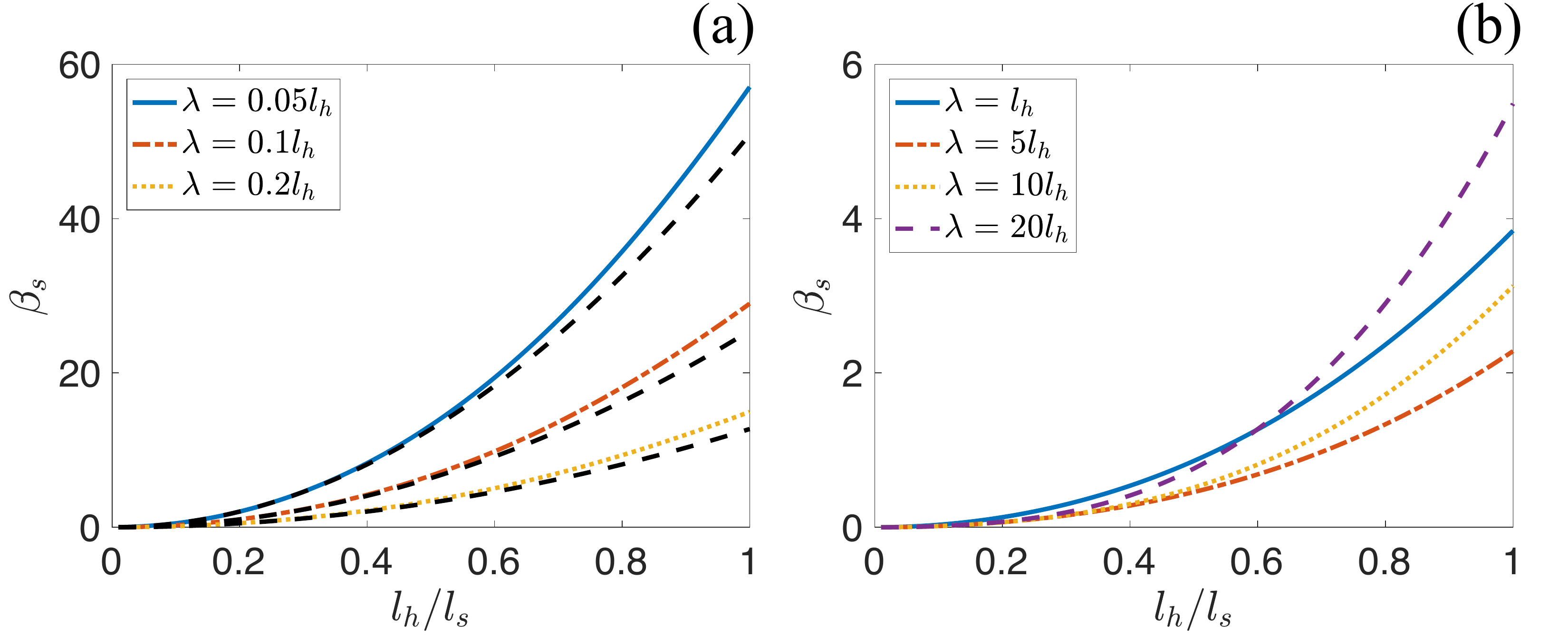}
\caption{\label{fig:beta} Adiabaticity parameter $\beta_s$ as a function of the inverse spin-relaxation length $\ell_s$ for different domain wall sizes. The results are plotted for $P_I=0.5$ and $k_Il_h=0.5$. The anlaytical estimate for $\beta_s$ in Eq.~\eqref{eq:beta} is shown as the black dashed curve in (a).}
\end{figure}

In order to get a further insight on the relative magnitudes of the force and torque, we examine the adiabaticity parameter $\beta_s=\lambda F/T_z$ as a function of $l_h/\ell_s$ for different $\lambda$ in Fig.~\ref{fig:beta}. Since $F$ and $T_z$ both decay in the same manner, $\beta_s$ is independent of the distance $X$ from the injector. Comparing the values of $\beta_s$ in Fig.~\ref{fig:beta}(a) and (b), we can see that $\beta_s$ is indeed smaller for larger domain walls, but the spin relaxation also plays an important role. We can see that $\beta_s\gg1$ for strong spin relaxation, i.e., force is much larger than the torque.
On the other hand, the torque is much larger than the force for large domain walls  $\lambda \gtrsim l_h$, provided the spin relaxation length is also longer than $l_h$ [Fig.~\ref{fig:beta}(b)]. 
For small domain walls $\lambda\ll l_h$, $\beta_{s}$ is proportional to $\lambda^{-1}$. We can estimate $\beta_s$ in this limit for $l_h<\ell_s$ by
\begin{equation}\label{eq:beta}
\beta_s=\frac{8}{\pi}\frac{l_h}{\lambda}\frac{l_h^2}{\ell_s^2}.
\end{equation}
This is plotted in Fig.~\ref{fig:beta}(a) as the black dashed curve.

This behavior can be compared to the case of strong ferromagnets in the ballistic limit \cite{PhysRevLett.92.086601}. There the only non-adiabaticity (non-vanishing $\beta_s$) comes from the finite $\lambda_F/\lambda$. The spin diffusion equation employed here assumes that the Fermi wavelength $\lambda_F$ is much smaller than any other length scale. However, we see that in this case other length scales, such as $l_h$ and $\ell_s$ govern the behavior of the adiabaticity parameter.

\section{Domain wall dynamics\label{sec:Domainwalldynamics}}
In the absence of an external pinning and a negligible domain wall mass,\cite{hurst2019electron} the dynamic equations of domain wall motion are~\cite{PhysRevLett.92.086601,TATARA2008213}
\begin{equation}\label{eq:DomainWallEqn1}
\Dot{\phi}+\alpha_0\frac{\Dot{X}}{\lambda}=\frac{\lambda}{\hbar NS}F
\end{equation}
\begin{equation}\label{eq:DomainWallEqn2}
\Dot{X}-\alpha_0\lambda\Dot{\phi}=\frac{K_\perp\lambda}{2\hbar}S\sin(2\phi)+\frac{\lambda}{\hbar NS}T_z,
\end{equation}
where $\phi$ is the out-of-plane angle in Eq.~\eqref{eq:ExchangeField}, $\alpha_0$ is the Gilbert damping parameter of the local magnetization, $K_\perp$ is the perpendicular anisotropy energy, and $S$ is the size of the localized spin. Also, $N=2\lambda W /a_0^3$ is the number of spins in the domain wall, and $a_0$ is the lattice constant. The force and torque are given in 
Eq.~\eqref{eq:Force} and in Eq.~\eqref{eq:Torque}, respectively. 

The unit of $F$ and $T_z/\lambda$ is $h\gamma VN_0W$. In order to make the dynamic equations dimensionless, we multiply
$$
t_0=\frac{\hbar NS}{\lambda h\gamma VN_0W}=\frac{2\hbar S}{a_0^3N_0h\gamma V}
$$
to both sides of Eq.~\eqref{eq:DomainWallEqn1} and Eq.~\eqref{eq:DomainWallEqn2}, and after reorganizing the terms, write
\begin{equation}\label{eq:UnitlessDomainWalleqn1}
\frac{\dot{X}}{\lambda}=\frac{1}{1+\alpha_0^2}\left[\alpha_0f+\frac{\tau_z}{\lambda}+k_\perp\sin(2\phi) \right]
\end{equation}
\begin{equation}\label{eq:UnitlessDomainWalleqn2}
\dot{\phi}=\frac{1}{1+\alpha_0^2}\left[f-\alpha_0\frac{\tau_z}{\lambda}-\alpha_0k_\perp\sin(2\phi) \right].
\end{equation}
Here we defined
$$
f=-\frac{\pi}{\lambda\gamma VN_0}\int dx s_1^0
$$
$$
\tau_z=-\frac{1}{\gamma VN_0}\int dx s_2^0\sin\theta
$$
$$
k_\perp=\frac{K_\perp S^2}{a_0^3N_0h\gamma V}.
$$

\begin{figure}[t]
\centering
\includegraphics[width=1\linewidth]{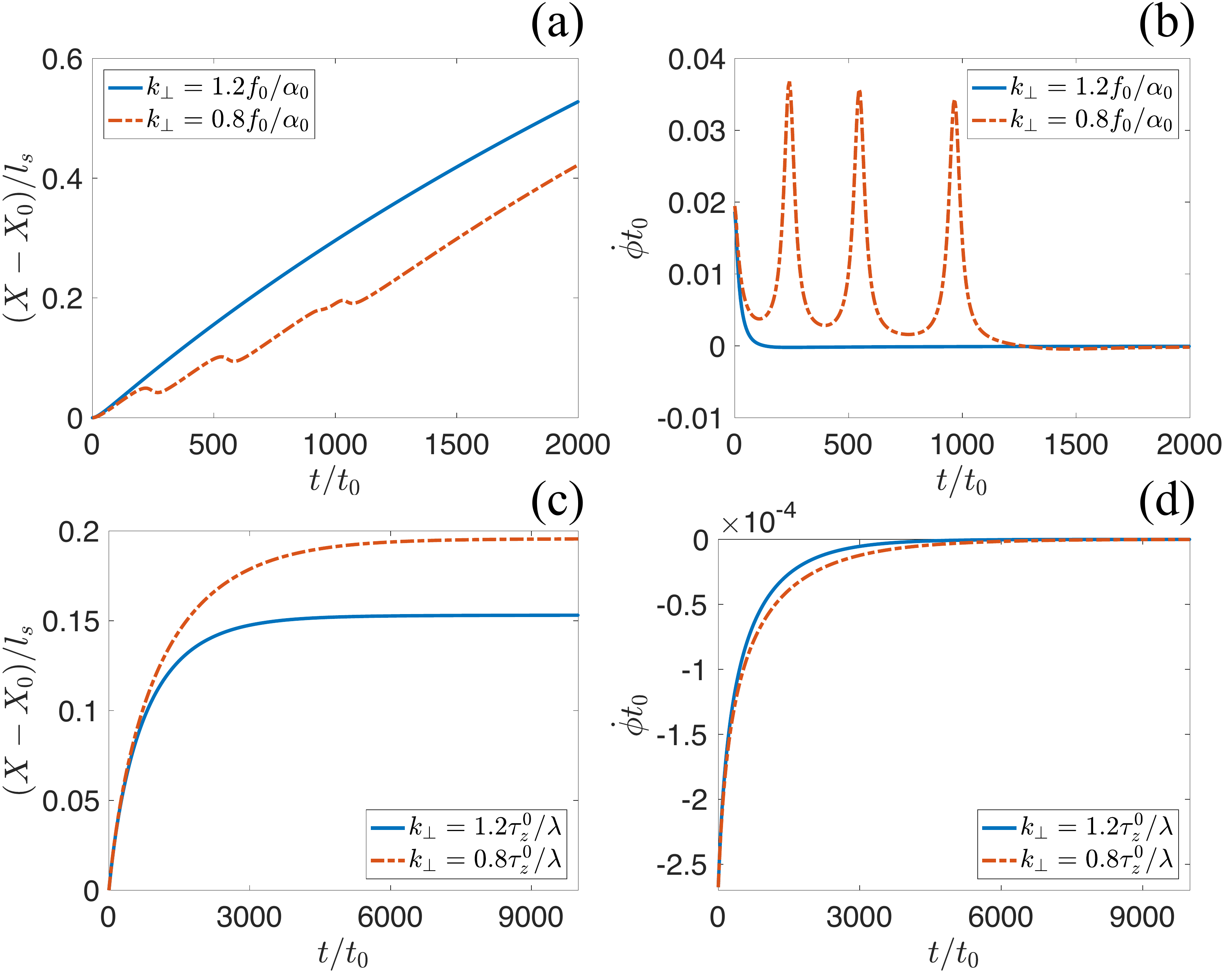}
\caption{\label{fig:dynamics1} Full numerical solutions of the dynamic equations of domain wall motion in Eq.~\eqref{eq:UnitlessDomainWalleqn1} and Eq.~\eqref{eq:UnitlessDomainWalleqn2}. The case where the force is much larger than the torque is shown in (a,b), and the one where the torque is much larger than the force is shown in (c,d). In (a,b) we use $\lambda=0.01l_h$ and $\ell_s=3.2 l_h$. In (c,d) $\lambda=20l_h$ and $\ell_s=100 l_h$. The other parameters used in the calculations are $P_I=0.5$, $k_Il_h=0.5$, $X_0=\lambda/2$, and $\alpha_0=0.2$. In the inset of (c,d), the results are shown for a smaller time scale. }
\end{figure}

We first discuss the case where the force is much larger than the torque ($\beta_s  \gtrsim 1$). We can see from Fig.~\ref{fig:beta} that this is the case for small domain walls and large domain walls with strong spin relaxation $l_h\gg \ell_s$. For convenience we consider a small domain wall $\lambda\ll l_h$. The full numerical solutions of the dynamic equations of domain wall motion in Eq.~\eqref{eq:UnitlessDomainWalleqn1} and Eq.~\eqref{eq:UnitlessDomainWalleqn2} are shown in Fig.~\ref{fig:dynamics1}. 

If the force is a constant $f=f_0$ in the absence of the torque, Eq.~\eqref{eq:UnitlessDomainWalleqn2} yields $\dot{\phi}=0$ for $f_0<\alpha_0k_\perp$. Then the domain wall moves with a constant velocity and a constant out-of-plane angle
\begin{equation}\label{eq:ConstantDomainWallVelocity}
\dot{X}=\frac{\lambda f_0}{\alpha_0},
\end{equation}
\begin{equation}\label{eq:PhiforConstantDomainWallVelocity}
\phi=\frac{1}{2}\arcsin\left(\frac{f_0}{\alpha_0 k_{\perp}}\right).
\end{equation}
In the spin current induced domain wall motion, the force decays as a function of the domain wall position $X$. If we write the force as $f=f_0e^{-X/\ell_s}$, then $\dot{\phi}\rightarrow0$ for $t\rightarrow\infty$, and this yields
\begin{equation}\label{eq:DomainWallEqnDrivenByForce}
\dot{X}=\frac{\lambda f_0}{\alpha_0}e^{-X/\ell_s}.
\end{equation}
This equation can be solved as
\begin{equation}
X=X(0)+\ell_s\log\left[1+\frac{f_0\lambda t}{\ell_s\alpha_0} \right],
\label{eq:Xvstime}
\end{equation}
and 
$$
\dot{X}=\frac{f_0\ell_s\lambda}{\ell_s\alpha_0+f_0\lambda t},
$$
where $X(0)$ is the domain wall position where $\dot{\phi}\rightarrow0$. This is exemplified by the curves in Fig.~\ref{fig:dynamics1}(a,b). There, the blue curve shows the behavior in the case where the force is everywhere below $\alpha_0 k_\perp$, and where $\dot \phi \rightarrow 0$ at around $t \approx 200 t_0$. 
From Eq.~\eqref{eq:UnitlessDomainWalleqn2}, we can also determine 
\begin{equation}\label{eq:PhiDecayingForce}
\phi=\frac{1}{2}\arcsin\left[\frac{f_0\ell_s}{k_{\perp}}\frac{e^{-X(0)/\ell_s}}{\ell_s\alpha_0+f_0\lambda t} \right].
\end{equation}

If $f_0>\alpha_0k_\perp$, the constant force leads to an oscillatory domain wall motion. This is known as the Walker breakdown.~\cite{doi:10.1063/1.1663252} The red dash-dotted curves in Fig.~\ref{fig:dynamics1}(a,b) shows the situation where the force is initially above this threshold, and only as the domain wall has moved further from the injector $f$ gets below this threshold (around $t \gtrsim 1500 t_0$). After that the domain wall motion follows Eq.~\eqref{eq:DomainWallEqnDrivenByForce}.

From Fig.~\ref{fig:beta}, we can see that the torque is much larger than the force for large domain walls and weak spin relaxation. In the case of a constant torque in the absence of the force, the domain wall does not move if $\tau_z^0<k_\perp\lambda$. The reason is that the perpendicular anisotropy energy described by the coefficient $k_\perp$ absorbs the torque completely. This is known as intrinsic pinning.~\cite{PhysRevLett.92.086601} Otherwise, if $\tau_z^0>k_\perp\lambda$, the domain wall moves with a finite velocity. Similar to the force, we can write the torque as $\tau_z=\tau_z^0e^{-X/\ell_s}$. When the torque decays until $\tau_z(X(t))<k_\perp\lambda$ so that $\dot{\phi}\rightarrow0$, the domain wall stops moving. It takes a longer time for a smaller $k_\perp$ to absorb the torque completely. The domain wall position and $\dot{\phi}$ as a function of time for a decaying torque are plotted in Fig.~\ref{fig:dynamics1}(c,d). 

\begin{figure}[t]
\centering
\includegraphics[width=1\linewidth]{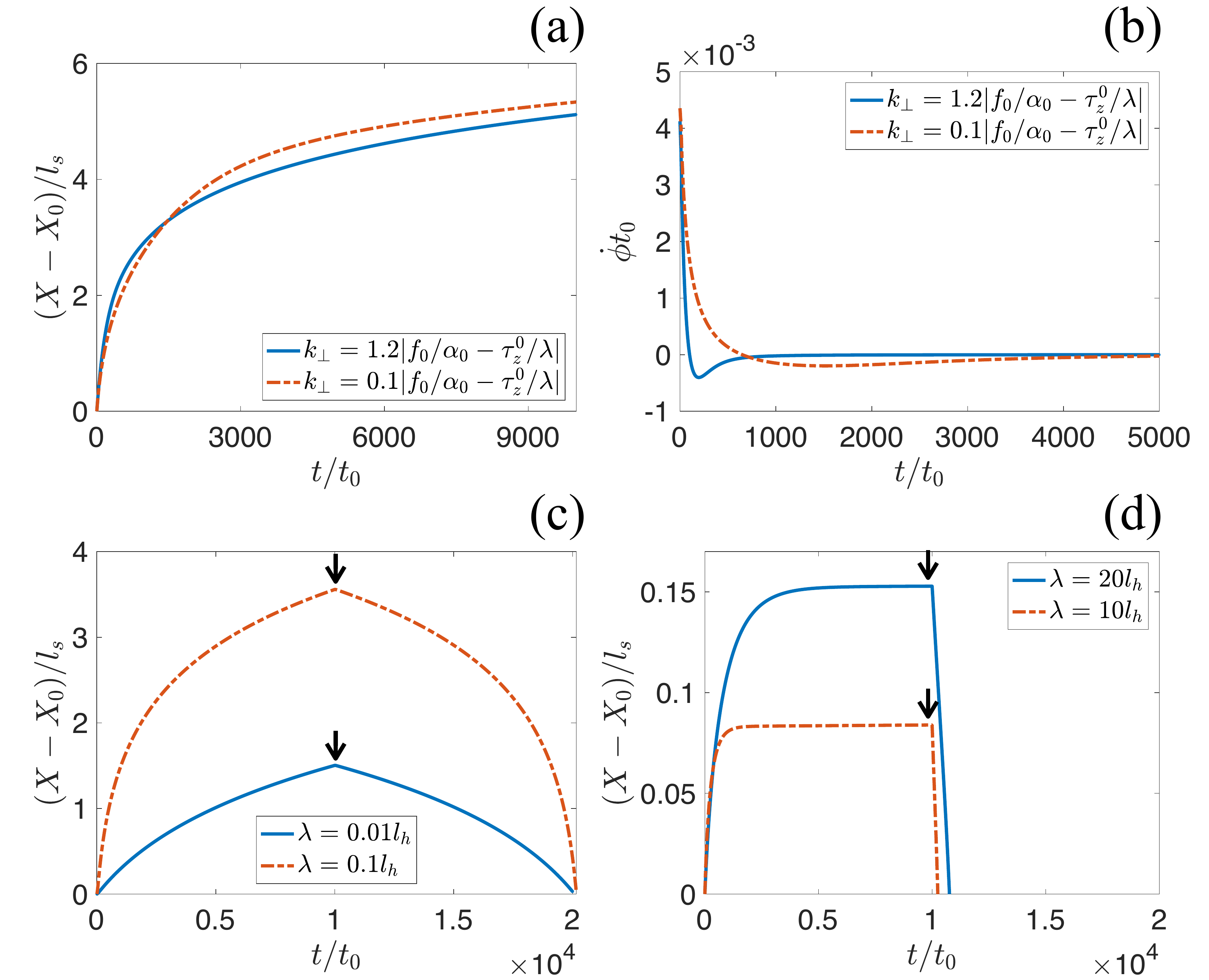}
\caption{\label{fig:dynamics2} Domain wall dynamics in the presence of comparable force and torque for different $k_{\perp}$ in (a,b). Forward and backward moving
domain walls
for different domain wall sizes in (c,d) for a fixed $k_\perp=1.2\vert f_0/\alpha_0-\tau_z^0/\lambda\vert$. In (a,b) we use $\lambda=l_h$ and $\ell_s=3.2l_h$. In (c) $\ell_s=3.2l_h$, and in (d) $\ell_s=100l_h$. The other parameters used in the calculations are $P_I=0.5$, $k_Il_h=0.5$, $X_0=\lambda/2$, and $\alpha_0=0.2$. In (c,d) we use a small arrow to denote when the voltage changes sign.}
\end{figure}

We next examine the domain wall motion in the presence of both force and torque ($\beta_s \approx 1$). In the case of constant force and torque, a small force is enough to destroy the intrinsic pinning. The domain wall moves with a constant velocity, see Eq.~\eqref{eq:ConstantDomainWallVelocity}. This is also the case with decaying force and torque, and the domain wall motion follows Eq.~\eqref{eq:DomainWallEqnDrivenByForce}. We can use Eq.~\eqref{eq:UnitlessDomainWalleqn2} to obtain $\phi$ for $\dot{\phi}\rightarrow0$ as
\begin{equation}\label{eq:PhiDecayingBoth}
\phi=\frac{1}{2}\arcsin\left[\frac{1}{k_{\perp}}\left(\frac{f_0}{\alpha_0}-\frac{\tau_z^0}{\lambda} \right)\frac{\alpha_0\ell_s}{\ell_s\alpha_0+f_0\lambda t}e^{-X(0)/\ell_s} \right].
\end{equation}
The numerical solutions of the dynamic equations of the domain wall motion in Eq.~\eqref{eq:UnitlessDomainWalleqn1} and Eq.~\eqref{eq:UnitlessDomainWalleqn2} in the presence of comparable force and torque are plotted in Fig.~\ref{fig:dynamics2}(a,b)

In the above discussions, the voltage at the injector is considered to be positive $V>0$. If the voltage changes sign at some instant of time, then the sign of the force and torque also changes, and they start pulling the domain wall instead of pushing it. This leads to the reversed motion of the domain wall. The reversed domain wall motion for small domain walls $\lambda<l_h$ are shown in Fig.~\ref{fig:dynamics2}(c). In this case, the domain wall reverses back to its original position $X=0$ at an equal amount of time as the one needed to push it further. The reversed domain wall motion for large domain walls with weak spin relaxation is shown in Fig.~\ref{fig:dynamics2}(d). In this case the domain wall had stopped before the sign change of the injected spin current.

The above analysis is based on the dynamics described by Eqs.~\eqref{eq:DomainWallEqn1} and \eqref{eq:DomainWallEqn2}, with force and torque obtained from the solutions of the spin diffusion equations. Those equations were derived \cite{TATARA2008213} by assuming a clean ferromagnet and an instant electronic response to the domain wall motion. It was shown\cite{PhysRevB.83.020410,wegrowe2012magnetization,olive2012beyond,bhattacharjee2012atomistic,olive2015deviation,PhysRevB.92.184410,PhysRevB.99.134409,hurst2019electron} that taking into account the delayed electron dynamics, extra "inertial" terms proportional to $\ddot \phi$ and $\ddot X$ can also appear, leading for example to a hysteretic dynamics of the domain wall. The prefactor of those terms, an effective mass of the domain wall, is proportional to the time it takes for the electrons to traverse the domain wall width $\lambda$. If $\lambda$ is large compared to the elastic mean free path, as assumed in the present manuscript, this effective mass is also likely to change from the ballistic limit considered in Ref.~\onlinecite{hurst2019electron}. This is why we did not yet consider its possible effect on the dynamics in the present manuscript.

\section{Domain wall  resistance\label{sec:InjectionResistance}}
The current induced from the injector electrode is given by [see Appendix \ref{sec:Current throughtheinjector}, Eq.~\eqref{eq:TunnellingCurrent}]
$$
I=G[-\Gamma \gamma V+P_Is_3(0)/N_0],
$$
where $G$ is the conductance of the injector, and $\Gamma$ is defined in Eq.~\eqref{eq:RescalingFactorkI}.
We can see from Appendix \ref{sec:SpinAccuInInhomoMagnt} that the spin accumulation is linear in the injection voltage $V$. Taking that into account allows us to include an extra resistance that depends on the relaxation of $s_3$ along the wire. In particular, we may study this extra resistance in the presence of the domain wall at position $X$, and without it (formally $X \rightarrow \infty$).
This domain wall resistance provides a direct method to detect the domain wall motion. 

\begin{figure}[ht]
\centering
\includegraphics[width=1\linewidth]{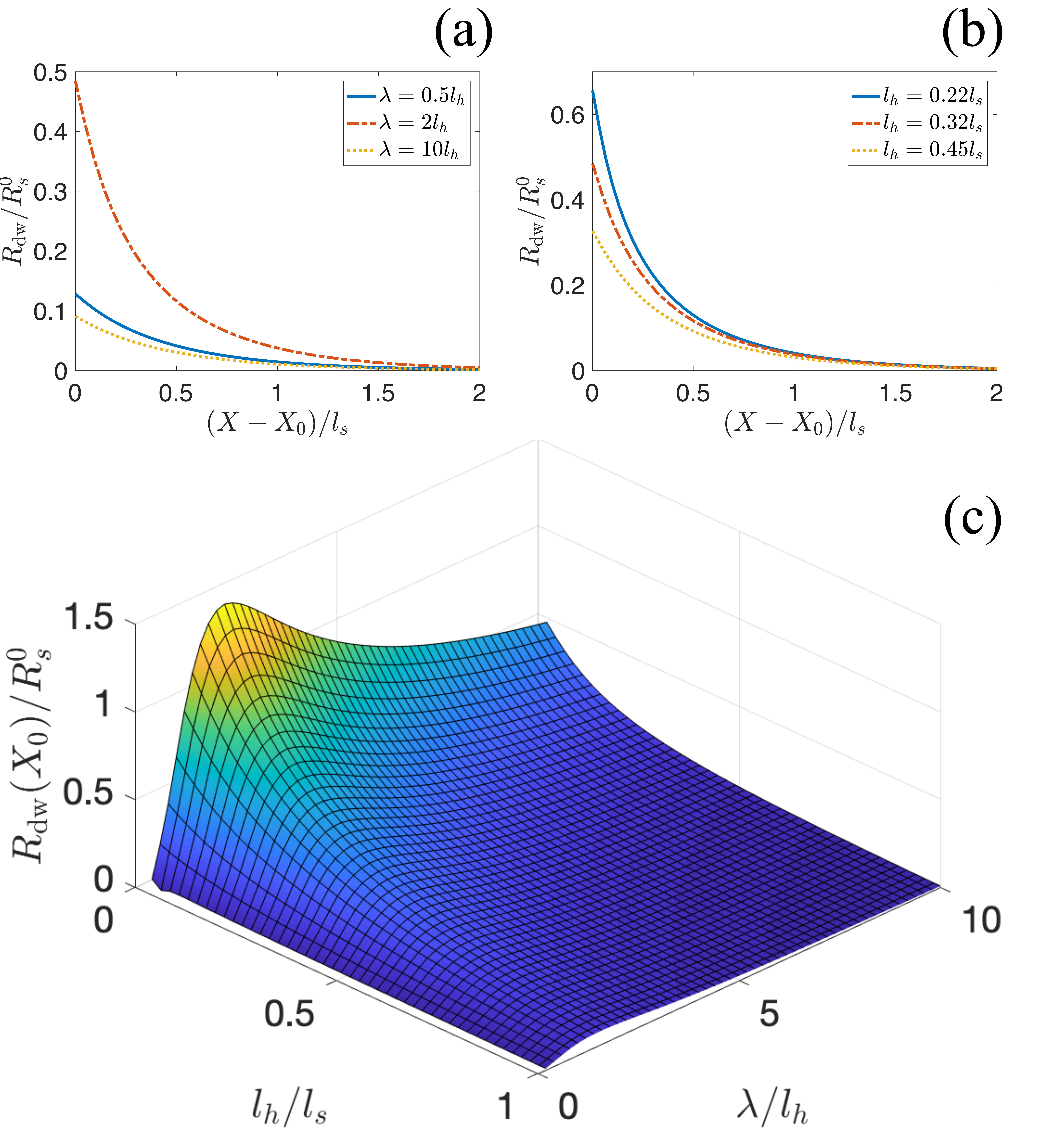}
\caption{\label{fig:resitance} The additional resistance $R_{\rm{dw}}$ introduced from the domain wall  plotted as a function of the domain wall position in (a,b), and the dependence of the maximum $R_{\rm{dw}}$ at $X_0$ plotted as a function of $\lambda$ and $\ell_s^{-1}$ in (c). In (a) $R_{\rm{dw}}$ is plotted for different domain wall sizes and $\ell_s=3.2 l_h$, and in (b) for different spin relaxation lengths for $\lambda=2l_h$. The other parameters used in the calculations are $k_Il_h=0.5$, $P_I=0.5$, and $X_0=\lambda/2$. }
\end{figure}

If we denote the spin accumulation at the position of the injector as $s_3(0)=\mu_zP_I\gamma VN_0$, where $\mu_z=\mu_z(X,l_h,\ell_s,k_I,\lambda)$ is a dimensionless quantity, then the current through the contact can be written as $I=G(-\Gamma+\mu_zP_I^2)\gamma V$. The spin accumulation thus adds a "spin resistance"
\begin{equation}\label{eq:SpinReistance}
R_s=\frac{1}{G\mu_zP_I^2\gamma}.
\end{equation}
The contribution of the domain wall to the spin resistance in Eq.~\eqref{eq:SpinReistance} can be found by 
taking the difference of $R_s$ with the resistance in the absence of the domain wall $R_s^0$ as $R_{\rm{dw}}=R_s-R_s^0$. Here 
$$
R_s^0=\frac{1}{G\mu_z(X\rightarrow\infty)P_I^2\gamma}=\frac{2+k_I\ell_s}{GP_I^2k_I\ell_s\gamma},
$$
where $\mu_z(X\rightarrow\infty)$ is determined from Eq.~\eqref{eq:C61Homo}, and $k_I$ is the injector transparency. Again the analytic formula for $R_{\rm{dw}}$ is long, but we show its behavior for some selected parameters in Fig.~\ref{fig:resitance}.

We can see that the domain wall contribution to the resistance $R_{\rm{dw}}$ reduces exponentially as the domain wall moves away from the injector, as is natural due to the fact that $R_{\rm{dw}}$ depends on the size of the spin accumulation around the domain wall. Close to the injector $X = \lambda/2$  
[Fig.~\ref{fig:resitance}(c)], the domain wall contribution is maximal for $l_h \ll \ell_s$ and for $\lambda \approx l_h$.

\section{Conclusion\label{sec:conclusion}}
In conclusion, we have studied the domain wall motion in weak ferromagnets in a non-local spin-injection setup. We have used a spin-diffusion equation to calculate the spin accumulation and evaluated the force and torque acting on the domain wall. Both decay exponentially as a function of domain wall position. We have studied the domain wall dynamics and have showed that the domain wall motion exhibits interesting features due to the decaying force and torque. For example, if the force close to the injector is larger than the torque and a threshold for Walker breakdown, the domain wall exhibits first an oscillatory dynamics, but further from the injector spin relaxation necessarily takes the force below that threshold value, resulting into an algebraically decaying domain wall speed. On the other hand, for a large torque close to the injector, compared to both the force and an intrinsic pinning value due to anisotropy, the relatively steady initial motion ceases when the torque becomes smaller than the intrinsic pinning value, and the domain wall essentially stops. Since the sign of both the force and the torque depend on the sign of the injection current, the domain wall motion can be reversed by reversing the sign of the current. This is why the pure spin current can also be used to pull the domain wall back towards the injector. Besides the analysis of the force and torque and their result on the dynamics, we have also described a means to detect the domain wall position via monitoring the injection resistance that depends on the domain wall position.

Our model is an alternative description of domain wall motion compared to majority of the models  \cite{TATARA2008213} dealing with essentially ballistic electron systems. In those cases the only relevant length scales are the domain wall size and the Fermi wavelength. We show how in disordered systems and weak ferromagnets there may be also other essential length scales governing the domain wall dynamics, especially the magnetic length $l_h$ and the spin relaxation length $\ell_s$. Our approach is made possible by the use of the spin diffusion equation also in the presence of inhomogeneous magnetism, which would not be straightforward when the spin polarization in the ferromagnet is large. To be able to use this equation, we hence need to assume weak ferromagnetism, which limits the applicability range of our approach. On the other hand, it provides hints on the types of effects expected also in the case of strong ferromagnets for which, to our knowledge, an analogous theory does not exist. 

Moreover, our model can be used as a reference in the study of the domain wall motion in superconductor/ferromagnetic insulator (S/FI) hybrid structures. In such structures, the spin accumulation in the superconductor is described by a set of kinetic equations, which reduce to the spin diffusion equation in the normal state. The equilibrium properties of such a structure are studied in Ref.~\onlinecite{PhysRevB.99.104504}. A thermally induced torque is studied in a superconductor/ferromagnet bilayer in the clean limit~\cite{bobkova2019thermally}, but to our knowledge this has not been extended to the experimentally more relevant disordered limit.

\begin{acknowledgments}
We thank M. Silaev for discussions. This work was supported by the Academy of Finland project number 317118.
\end{acknowledgments}

\appendix
\section{Spin diffusion equation\label{sec:DerivationSpinDiffusionEquation}}
As a useful tool in describing the electronic transport properties of magnetic materials, we start by the spin dependent Boltzmann equation in the diffusive limit~\cite{heikkila2013physics,PhysRevB.48.7099}
\begin{equation}\label{eq:BoltzmannEquationWithSpin}
(\partial_t - D\nabla^2)f_z(\boldsymbol{r},\epsilon,t)=-\frac{1}{\tau_{s}}f_z(\boldsymbol{r},\epsilon,t),
\end{equation}
where $D$ is the diffusion constant (in a weak ferromagnet assumed independent of the spin index), $f_z=f_{\uparrow}-f_{\downarrow}$ and $f_{\sigma}$ is the distribution function of electrons with spin $\sigma={\uparrow/\downarrow}$, and $\tau_{s}$ is the spin-flip relaxation time. This equation has been widely used in spintronics, for example in the description of the spin accumulation at an interface between a ferromagnet and a nonmagnetic metal.~\cite{PhysRevLett.58.2271}

Here we assume a quasistatic description, where the domain wall moves slowly compared to the electrons, so that the spin diffusion equation can be solved in the static case. We can estimate the validity range of this assumption by comparing the time scale $\tau=\ell_s \alpha_0/(f_0 \lambda$) of domain wall motion through a length $\ell_s$ (given in Eq.~\eqref{eq:Xvstime}) to the characteristic electron diffusion time $D/\ell_s^2$ through a similar length scale. The quasistatic approximation is valid when $\tau \gg D/\ell_s^2$, or
\begin{equation}
    eV \ll \frac{\hbar D}{\lambda \ell_s} \frac{2 S \alpha_0}{N_0 a_0^3 h \gamma},
\end{equation}
where we assumed $f_0$ of the order of unity. In this limit we can hence disregard the time derivative in Eq.~\eqref{eq:BoltzmannEquationWithSpin}.

In the case of an inhomogeneous exchange field, other spin components should be taken into account, and we can replace $f_s$ by $\boldsymbol{f}\cdot\boldsymbol{\sigma}=(f_x,f_y,f_z)\cdot\boldsymbol{\sigma}$, where $\boldsymbol{\sigma}=(\sigma_1,\sigma_2,\sigma_3)$ is a vector of Pauli spin matrices. Considering the Heisenberg equation of motion for $\boldsymbol{f}\cdot\boldsymbol{\sigma}$, and substituting back to the Boltzmann equation (see a similar derivation in Refs.~\onlinecite{PhysRevB.62.5700,brataas2006non}, except that those articles write an opposite sign of the Zeeman energy term) we obtain for a steady state, 
$$
D\nabla^2\boldsymbol{f}\cdot\boldsymbol{\sigma}=\frac{1}{\tau_{s}}\boldsymbol{f}\cdot\boldsymbol{\sigma}+\frac{i}{\hbar}\left[\frac{g\mu_B}{2}\boldsymbol{B}\cdot\boldsymbol{\sigma},\boldsymbol{f}\cdot\boldsymbol{\sigma} \right],
$$
where the other component of the commutator is the Zeeman energy. There, $g=2$ is the $g$-factor, $\mu_B$ is the Bohr magneton, and $\boldsymbol{B}$ is the magnetic field. By denoting $\boldsymbol{h}=g\mu_B\boldsymbol{B}/2$ and reorganizing the terms, we obtain
$$
\hbar D\nabla^2\boldsymbol{f}=\frac{\hbar}{\tau_{s}}\boldsymbol{f}-2\boldsymbol{h}\times\boldsymbol{f},
$$
where we used the relation $(\boldsymbol{a}\cdot\boldsymbol{\sigma})(\boldsymbol{b}\cdot\boldsymbol{\sigma})=2i(\boldsymbol{a}\times\boldsymbol{b})\cdot\boldsymbol{\sigma}$. This equation is an extension of Eq.~\eqref{eq:BoltzmannEquationWithSpin} to the case with inhomogeneous magnetization, as it reduces to Eq.~\eqref{eq:BoltzmannEquationWithSpin} for the case of homogeneous magnetization in the steady state.

Integrating over energy on the two sides, we finally obtain the spin-diffusion equation
\begin{equation}\label{eq:SpinDiffusionEquation}
\hbar D\nabla^2\boldsymbol{s}=\frac{\hbar}{\tau_{s}}\boldsymbol{s}-2\boldsymbol{h}\times\boldsymbol{s},
\end{equation}
where 
$$
\boldsymbol{s}(\boldsymbol{r})=N_0 \int d\epsilon \boldsymbol{f}(\boldsymbol{r},\epsilon)
$$
is the spin accumulation at position $\boldsymbol{r}$. The spin diffusion equation was used to describe the spin Hanle effect in ferromagnet-normal metal-ferromagnet systems.\cite{PhysRevB.62.5700,jedema2001electrical,PhysRevB.67.085319} Here we use it to calculate the spin accumulation in a weak ferromagnet, including the Hanle effect from the inhomogeneous exchange field.

The spin current is given by the derivative of the spin accumulation
$$
\boldsymbol{j}(\boldsymbol{r})=\hbar D\nabla\boldsymbol{s}(\boldsymbol{r}). 
$$
The spin current is a tensor, as it depends on position for all three spin components. This spin current plays an important role in the domain wall motion. 

\section{Spin accumulation with inhomogenous magnetization\label{sec:SpinAccuInInhomoMagnt}}
Since the rotation angle in Eq.~\eqref{eq:RotationAngle} is a step function, the spin diffusion equation in Eq.~\eqref{eq:SpinDiffEqnUnrotated} is separated into three regions. On the left and right side of the domain wall, the general solution of Eq.~\eqref{eq:SpinDiffEqnUnrotated} is given by
\begin{widetext}
$$
s_1^0=\frac{1}{\eta}\{ \cosh(z\eta\nu)\left[\eta C_{1i}\cos(z\eta\mu)+(\mu C_{2i}+\nu C_{4i})\sin(z\eta\mu) \right]+\left[(\nu C_{2i}-\mu C_{4i})\cos(z\eta\mu)+\eta C_{3i}\sin(z\eta\mu) \right]\sinh(z\eta\nu) \}
$$
$$
s_2^0=\frac{1}{\eta}\{ \cosh(z\eta\nu)\left[\eta C_{3i}\cos(z\eta\mu)+(-\nu C_{2i}+\mu C_{4i})\sin(z\eta\mu) \right]+\left[(\mu C_{2i}+\nu C_{4i})\cos(z\eta\mu)-\eta C_{1i}\sin(z\eta\mu) \right]\sinh(z\eta\nu) \}
$$
$$
s_3^0=e^{z/\ell_s}C_{5i}+e^{-z/\ell_s}C_{6i},
$$
where $i=1,3$ refers to the left and right side of the domain wall, and $C_{ni}$ are constants which are determined from the boundary conditions. Here we also defined
$$
\eta=\left(\frac{4}{l_h^4}+\frac{1}{\ell_s^4}\right)^{1/4}
$$
$$
\mu=\sin\left[\frac{1}{2}\arctan{\left( \frac{2\ell_s^2}{l_h^2}\right)} \right]
$$
$$
\nu=\cos\left[\frac{1}{2}\arctan{\left( \frac{2\ell_s^2}{l_h^2}\right)} \right].
$$

In the domain wall region the solutions are given by
\begin{equation*}
\begin{split}
s_1^0=&-\frac{C_{12}e^{zk_1}(\alpha^2+\ell_s^{-2}-k_1^2)}{2\alpha\sqrt{N_1}k_1}+\frac{C_{22}e^{-zk_1}(\alpha^2+\ell_s^{-2}-k_1^2)}{2\alpha\sqrt{N_1}k_1}-\frac{C_{32}e^{zk_2}(\alpha^2+\ell_s^{-2}-k_2^2)}{2\alpha\sqrt{N_2}k_2}\\
&+\frac{C_{42}e^{-zk_2}(\alpha^2+\ell_s^{-2}-k_2^2)}{2\alpha\sqrt{N_2}k_2}-\frac{C_{52}e^{zk_2^*}(\alpha^2+\ell_s^{-2}-k_2^{*2})}{2\alpha\sqrt{N_2}k_2^*}+\frac{C_{62}e^{-zk_2^*}(\alpha^2+\ell_s^{-2}-k_2^{*2})}{2\alpha\sqrt{N_2}k_2^*}
\end{split}
\end{equation*}
\begin{equation*}
s_2^0=\frac{C_{12}e^{zk_1}al_h^2}{36\alpha\beta^2\sqrt{N_1}k_1}-\frac{C_{22}e^{-zk_1}al_h^2}{36\alpha\beta^2\sqrt{N_1}k_1}+\frac{C_{32}e^{zk_2}bl_h^2}{72\alpha\beta^2\sqrt{N_2}k_2}-\frac{C_{42}e^{-zk_2}bl_h^2}{72\alpha\beta^2\sqrt{N_2}k_2}+\frac{C_{52}e^{zk_2^*}b^*l_h^2}{72\alpha\beta^2\sqrt{N_2}k_2^*}-\frac{C_{62}e^{-zk_2^*}b^*l_h^2}{72\alpha\beta^2\sqrt{N_2}k_2^*}
\end{equation*}
\begin{equation*}
s_3^0=\frac{C_{12}e^{zk_1}}{\sqrt{N_1}}+\frac{C_{22}e^{-zk_1}}{\sqrt{N_1}}+\frac{C_{32}e^{zk_2}}{\sqrt{N_2}}+\frac{C_{42}e^{-zk_2}}{\sqrt{N_2}}+\frac{C_{52}e^{zk_2^*}}{\sqrt{N_2}}+\frac{C_{62}e^{-zk_2^*}}{\sqrt{N_2}},
\end{equation*}
where 
$$
\alpha=\frac{\pi}{\lambda}
$$
$$
\beta=\left[\alpha^6+\frac{90\alpha^2}{l_h^4}+\frac{36\alpha^4}{\ell_s^2}+\frac{1}{2}\sqrt{-4\left(\alpha^4-\frac{12}{l_h^4}-\frac{12\alpha^2}{\ell_s^2} \right)^3+4\left(\alpha^6+\frac{90\alpha^2}{l_h^4}+\frac{36\alpha^4}{\ell_s^2} \right)^2} \right]^{1/3}
$$
and
$$
k_1=\sqrt{\frac{1}{3}\left(-2\alpha^2+\beta+\frac{3}{\ell_s^2}+\frac{\alpha^4-12/l_h^4-12\alpha^2/\ell_s^2}{\beta} \right)}
$$
$$
k_2=\sqrt{\frac{1}{12}\left[ -8\alpha^2+2i(i+\sqrt{3})\beta+\frac{12}{\ell_s^2}+\frac{2(1+i\sqrt{3})(-\alpha^4+12/l_h^4+12\alpha^2/\ell_s^2)}{\beta} \right]}
$$
are the solutions of the following characteristic equation
$$
\alpha^4\left(k^2-\frac{1}{\ell_s^2}\right)+\left[\frac{4}{l_h^4}+\left(k^2-\frac{1}{\ell_s^2} \right)^2 \right]\left(k^2-\frac{1}{\ell_s^2} \right)-2\alpha^2\left(\frac{2}{l_h^4}-k^4+\frac{1}{\ell_s^4} \right)=0.
$$
The other coefficients are
$$
a=\left(\alpha^4+\alpha^2\beta+\beta^2-\frac{12}{l_h^4} \right)^2+\frac{12\alpha^2}{\ell_s^2}\left(-2\alpha^4-2\alpha^2\beta+\beta^2+\frac{24}{l_h^4} \right)+\frac{144\alpha^4}{\ell_s^4}
$$
\begin{equation*}
\begin{split}
b=&-(1-i\sqrt{3})\alpha^8-(1+i\sqrt{3})\beta^4-\frac{48\beta^2}{l_h^4}-\frac{144(1-i\sqrt{3})}{l_h^8}+\alpha^6\left[-2\left(1+i\sqrt{3} \right)\beta+\frac{24}{\ell_s^2}\left(1-i\sqrt{3} \right) \right]\\
&+2\alpha^2\left[-(1-i\sqrt{3})\beta^3+\frac{12(1+i\sqrt{3})\beta}{l_h^4}+\frac{12\beta^2}{\ell_s^2}-\frac{144(1-i\sqrt{3})}{l_h^4\ell_s^2} \right]\\
&+6\alpha^4\left[ \beta^2+\frac{4(1+i\sqrt{3})\beta}{\ell_s^2}+4(1-i\sqrt{3})\left( \frac{1}{l_h^4}-\frac{6}{\ell_s^4} \right) \right]
\end{split}
\end{equation*}
$$
N_1=\frac{(1+l_h^2k_1^2)\left\lbrace a^2l_h^4+324\beta^4\left[\alpha^4-(k_1^2-\ell_s^{-2})^2+2\alpha^2(k_1^2+\ell_s^{-2}) \right] \right\rbrace}{1296\alpha^2\beta^4k_1^2}
$$
$$
N_2=\frac{(1+l_h^2\vert k_2\vert^2)\left\lbrace \vert b\vert^2l_h^4+1296\beta^4\left[\alpha^4+\vert k_2^2\vert^2+\ell_s^{-4}-2\ell_s^{-2}{\rm{Re}}(k_2^2)+\alpha^2(2\ell_s^{-2}-2{\rm{Re}}(k_2^2)+4\vert k_2\vert^2) \right] \right\rbrace}{5184\alpha^2\beta^4\vert k_2\vert^2 }.
$$

The unknown coefficients are determined from the boundary conditions. However, these coefficients are too long to be printed here and rather have to be shown numerically.\cite{mathematicanote} We plot some of the coefficients for different values of the domain wall size $\lambda$ in Fig.~\ref{fig:coefficients}.
\begin{figure}[t]
\centering
\includegraphics[width=0.8\linewidth]{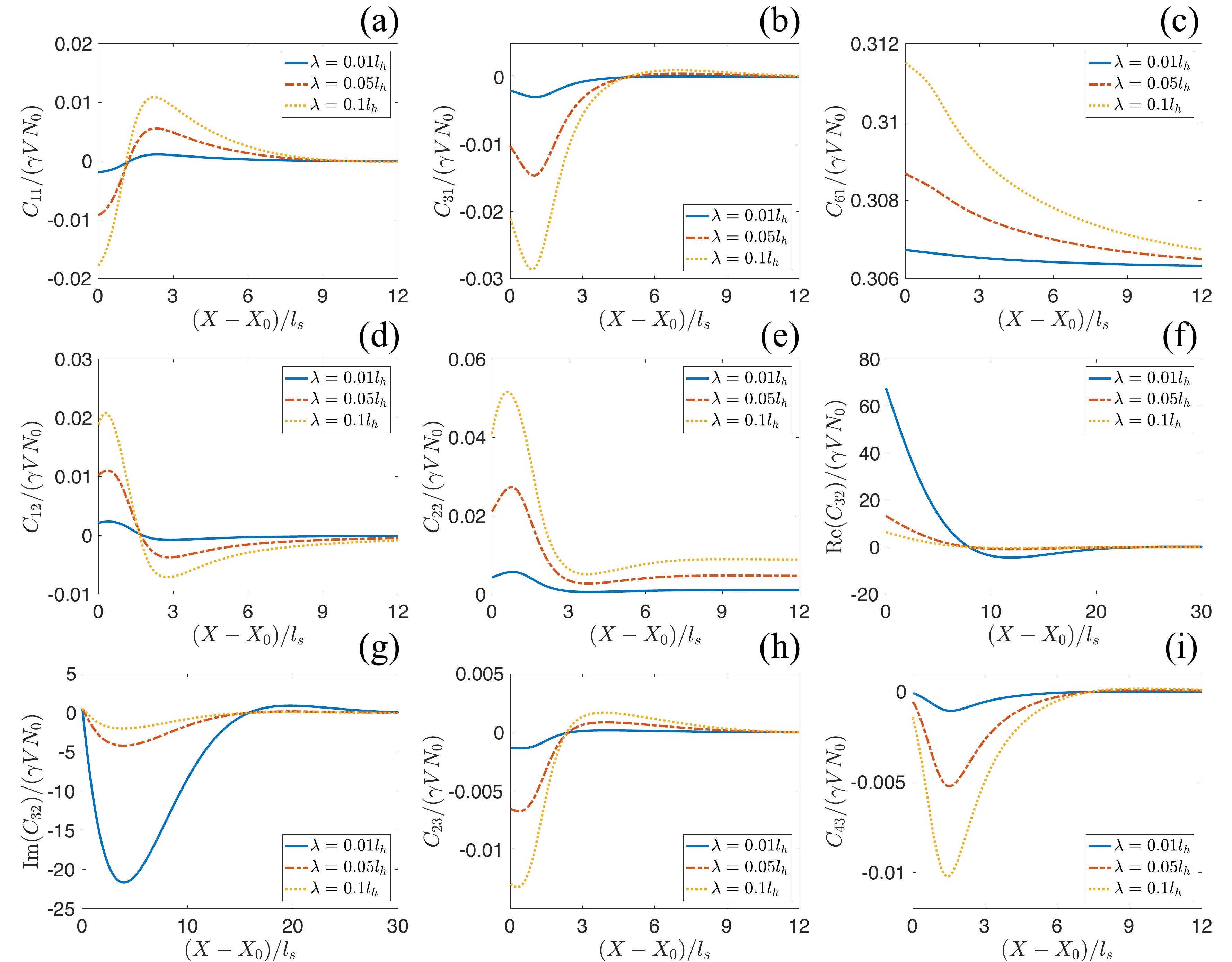}
\caption{\label{fig:coefficients}  Coefficients in the general solutions of the spin diffusion equation. Here the results are plotted for $\ell_s=3.2l_h$, $P_I=0.5$, $k_Il_h=0.5$, and $X_0=\lambda/2$.}
\end{figure}

The solutions also yield $C_{52}=C_{32}^*$ and $C_{62}=C_{42}^*$, which also imply real-valued spin accumulation. Moreover, the coefficients in region $i=3$ are very similar with those in region $i=1$, but with opposite signs ($C_{13}$, $C_{33}$ and $C_{63}$). For  $X\gg \ell_s$, we also find 
\begin{equation}\label{eq:C61Homo}
C_{61}=-C_{63}=\frac{k_I\ell_sP_I\gamma VN_0}{2+k_I\ell_s},
\end{equation}
but in general the expression is more complicated.

The solutions in the domain wall region can hence be written as
\begin{equation}\label{eq:AnalSpinAccumulationRottated1}
s_1^0=-\frac{\alpha^2+\ell_s^{-2}-k_1^2}{2\alpha\sqrt{N_1}k_1}\left(C_{12}e^{zk_1}-C_{22}e^{-zk_1}\right)-{\rm{Re}}\left[\frac{\alpha^2+\ell_s^{-2}-k_2^2}{\alpha\sqrt{N_2}k_2}\left(C_{32}e^{zk_2}-C_{42}e^{-zk_2} \right) \right]
\end{equation}
\begin{equation}\label{eq:AnalSpinAccumulationRottated2}
s_2^0=\frac{al_h^2}{36\alpha\beta^2\sqrt{N_1}k_1}\left(C_{12}e^{zk_1}-C_{22}e^{-zk_1} \right)+{\rm{Re}}\left[\frac{bl_h^2}{36\alpha\beta^2\sqrt{N_2}k_2}\left(C_{32}e^{zk_2}-C_{42}e^{-zk_2}\right) \right]
\end{equation}
\begin{equation}\label{eq:AnalSpinAccumulationRottated3}
s_3^0=\frac{1}{\sqrt{N_1}}\left(C_{12}e^{zk_1}+C_{22}e^{-zk_1} \right)+\frac{2}{\sqrt{N_2}}{\rm{Re}}\left(C_{32}e^{zk_2}+C_{42}e^{-zk_2} \right).
\end{equation}

The unrotated spin accumulation is given by the transformation in Eq.~\eqref{eq:UnrotSpinAccumulation}. We use these results to calculate the force and torque in the equations of domain wall motion. 

\end{widetext}

\section{Description of the injector\label{sec:Current throughtheinjector}}
We consider the case where the spin polarized current is injected to the weak ferromagnet from a strong ferromagnet attached to it a position $z=0$.
Since we assume the injector magnetization to be homogeneous, we can write the spin diffusion equation separately in the two spin directions collinear with the magnetization of the strong ferromagnet. If we assume that the current is injected into the weak ferromagnet from a wire placed in the $y$ direction, the spin-diffusion equation in the injector becomes \cite{brataas2006non} 
$$
\partial_y^2 s_{\sigma}^I=\frac{s_{\sigma}^I-s_{\Bar{{\sigma}}}^I}{2l_{\sigma}^2},
$$
where
$\Bar{{\sigma}}$ is the opposite spin to ${\sigma}=\uparrow/\downarrow$
and $l_{\sigma}=\sqrt{D_{\sigma}\tau_{\sigma}}$ is the spin-dependent spin relaxation length. The general solution of this equation can be written as
\begin{equation*}
\begin{split}
&s_{\uparrow/\downarrow}^I=\frac{l_{\downarrow}^2(C_1+C_2y)+l_{\uparrow}^2(C_3+C_4y)}{l_{\rm tot}^2}\\
&\pm \frac{l_{\downarrow/\uparrow}^2}{l_{\rm tot}^2}\left[(C_3-C_1)\cosh\left(\frac{y}{l} \right)+l(C_4-C_2)\sinh\left(\frac{y}{l} \right) \right],
\end{split}
\end{equation*}
where $l_{\rm tot}^2=l_{\uparrow}^2+l_{\downarrow}^2$ and $l=\sqrt{2}l_{\uparrow}l_{\downarrow}/l_{\rm tot}$. The unknown coefficients can be determined from the boundary condition~\cite{heikkila2013physics}
\begin{equation}\label{eq:BoundaryConditionInTheInjector}
\sigma_{\uparrow/\downarrow}^I A_T\partial_y s_{\uparrow/\downarrow}^I(0)=\frac{1}{R_I}\left[s_{\uparrow/\downarrow}^w(0)-s_{\uparrow/\downarrow}^I(0) \right],
\end{equation}
where $\sigma_{\sigma}^I=e^2N_{\sigma}D_{\sigma}$ is the spin dependent conductivity in the injector, $N_{\sigma}$ is the density of states of spin ${\sigma}$ at the Fermi level, $A_T$ is the cross-sectional area of the tunnelling junction, $R_I$ is the resistance of the contact between the injector and the wire, and $s_{\sigma}^w$ is the spin density for spin ${\sigma}$ created at the weak ferromagnet. If the voltage is applied at a distance $L$ away from the contact, then we have two more boundary conditions
\begin{equation*}
s_{\uparrow}^I(-L)+s_{\downarrow}^I(-L)=VN_0
\end{equation*}
$$
s_{\uparrow}^I(-L)-s_{\downarrow}^I(-L)=0,
$$
where the upper equation states that the average potential of the electrons at the distance $L$ is $V$ ($e=1$), and the lower indicates the vanishing of the spin accumulation in the electrode where the voltage is applied.

\begin{widetext}
With the determined coefficients, we can write the chemical potential and the spin accumulation at the position of injection as
\begin{equation}\label{eq:TotalSpinDensityInInjector}
\begin{split}
\mu_I(0)N_0=&s_{\uparrow}^I(0)+s_{\downarrow}^I(0)\\
=&\frac{l_{\rm{tot}}^2VN_0\sigma_{\uparrow}\sigma_{\downarrow}+2a_IL\left[l_{\downarrow}^2\sigma_{\uparrow}s_{\downarrow}^w(0)+l_{\uparrow}^2\sigma_{\downarrow}s_{\uparrow}^w(0)\right]}{l_{\rm{tot}}^2\sigma_{\uparrow}\sigma_{\downarrow}+a_IL(l_{\uparrow}^2\sigma_{\downarrow}+l_{\downarrow}^2\sigma_{\uparrow})+a_Il(a_ILl_{\rm{tot}}^2+l_{\downarrow}^2\sigma_{\downarrow}+l_{\uparrow}^2\sigma_{\uparrow})\tanh{(L/l)}}\\
&+\frac{\left\lbrace a_ILl_{\rm{tot}}^2\mu_w(0)N_0+l_{\rm{tot}}^2VN_0\sigma_F+(l_{\uparrow}^2-l_{\downarrow}^2)\left[\sigma_{\uparrow}s_{\downarrow}^w(0)-\sigma_{\downarrow}s_{\uparrow}^w(0) \right]\right\rbrace a_Il\tanh(L/l)}{l_{\rm{tot}}^2\sigma_{\uparrow}\sigma_{\downarrow}+a_IL(l_{\uparrow}^2\sigma_{\downarrow}+l_{\downarrow}^2\sigma_{\uparrow})+a_Il(a_ILl_{\rm{tot}}^2+l_{\downarrow}^2\sigma_{\downarrow}+l_{\uparrow}^2\sigma_{\uparrow})\tanh{(L/l)}}
\end{split}
\end{equation}
\begin{equation}\label{eq:DifferenceSpinDensityInInjector}
s_3^I(0)=s_{\uparrow}^I(0)-s_{\downarrow}^I(0)=\frac{a_Ill_{\rm{tot}}^2\left[a_ILs_3^w(0)+VN_0(\sigma_{\uparrow}-\sigma_{\downarrow})/2-\sigma_{\uparrow}s_{\downarrow}^w(0)+\sigma_{\downarrow}s_{\uparrow}^w(0) \right]\tanh(L/l)}{l_{\rm{tot}}^2\sigma_{\uparrow}\sigma_{\downarrow}+a_IL(l_{\uparrow}^2\sigma_{\downarrow}+l_{\downarrow}^2\sigma_{\uparrow})+a_Il(a_ILl_{\rm{tot}}^2+l_{\downarrow}^2\sigma_{\downarrow}+l_{\uparrow}^2\sigma_{\uparrow})\tanh{(L/l)}},
\end{equation}
where $a_I=1/(R_IA_T)$ and $\sigma_F^I=(\sigma_{\uparrow}^I+\sigma_{\downarrow}^I)/2$. Here we also defined the chemical potential and the spin accumulation in the weak ferromagnet as $\mu_w(0)N_0=s_{\uparrow}^w(0)+s_{\downarrow}^w(0)$ and $s_3^w(0)=s_{\uparrow}^w(0)-s_{\downarrow}^w(0)$. 

We assume for simplicity that the injector and the weak ferromagnetic wire cross sections are equal. Defining the injector transparency $\kappa_I=1/(\sigma^wR_IA_T)$, 
we can write the boundary condition analogous to Eq.~\eqref{eq:BoundaryConditionInTheInjector} for the weak ferromagnet wire as
$$
\partial_z s_{\uparrow/\downarrow}^w(0)=\kappa_I\left[s_{\uparrow/\downarrow}^w(0)-s_{\uparrow/\downarrow}^I(0) \right],
$$
where $\sigma^w$ is the conductivity in the weak ferromagnet.
We then write this boundary condition in terms of $\mu_w(0)$ and $s_3^w(0)$, and choose the zero point of potential so that $\mu_w(0)=0$. By substituting $s_{\uparrow/\downarrow}^I(0)$ in Eq.~\eqref{eq:TotalSpinDensityInInjector} and Eq.~\eqref{eq:DifferenceSpinDensityInInjector}, we obtain for $l_{\uparrow}=l_{\downarrow}$
$$
\begin{pmatrix}
\partial_z \mu_w(0)N_0\\
\partial_z s_3^w(0)
\end{pmatrix}=\frac{\kappa_I}{a_Il\tanh(L/l)(a_IL+\sigma_F^I)+a_IL\sigma_F^I+\sigma_{\uparrow}^I\sigma_{\downarrow}^I}
\begin{pmatrix}
a_Il\tanh(L/l)\sigma_F^I+\sigma_{\uparrow}^I\sigma_{\downarrow}^I & a_IL(\sigma_{\uparrow}^I-\sigma_{\downarrow}^I)/2\\
a_Il\tanh(L/l)(\sigma_{\uparrow}^I-\sigma_{\downarrow}^I)/2 & a_IL\sigma_F^I+\sigma_{\uparrow}^I\sigma_{\downarrow}^I
\end{pmatrix}
\begin{pmatrix}
-VN_0\\
s_3^w(0)
\end{pmatrix}.
$$

\end{widetext}

This equation leads to an Onsager relation for the current through the contact
\begin{equation}\label{eq:OnsagerRelation}
\begin{pmatrix}
\partial_z \mu_w(0)N_0\\
\partial_z s_3^w(0)
\end{pmatrix}=
\begin{pmatrix}
\Gamma k_I & P_Ik_I\\
P_Ik_I & k_I
\end{pmatrix}
\begin{pmatrix}
-\gamma VN_0\\
s_3^w(0)
\end{pmatrix},
\end{equation}
where the injector polarization and transparency are defined as
\begin{equation}
P_I=\frac{a_IL(\sigma_{\uparrow}^I-\sigma_{\downarrow}^I)/2}{a_IL\sigma_F^I+\sigma_{\uparrow}^I\sigma_{\downarrow}^I}=\frac{L(\sigma_{\uparrow}^I-\sigma_{\downarrow}^I)}{2(L\sigma_F^I+\sigma_{\uparrow}^I\sigma_{\downarrow}^IR_IA_T)}
\end{equation}
\begin{equation}
\begin{split}
&k_I = \frac{\kappa_I(a_IL\sigma_F^I+\sigma_{\uparrow}^I\sigma_{\downarrow}^I)}{a_Il\tanh(L/l)(a_IL+\sigma_F^I)+a_IL\sigma_F^I+\sigma_{\uparrow}^I\sigma_{\downarrow}^I}\\
&=\frac{(L\sigma_F^I+\sigma_{\uparrow}^I\sigma_{\downarrow}^IR_IA_T)/\sigma^w}{l\tanh(L/l)(L+R_IA_T\sigma_F^I)+L\sigma_F^IR_IA_T+\sigma_{\uparrow}^I\sigma_{\downarrow}^IR_I^2A_T^2},
\end{split}
\end{equation}
\begin{equation}\label{eq:RescalingFactorVoltage}
\gamma = \frac{l}{L}\tanh\left(\frac{L}{l} \right)
\end{equation}
and
\begin{equation}\label{eq:RescalingFactorkI}
\begin{split}
\Gamma&=\frac{L}{l}\frac{a_Il\tanh(L/l)\sigma_F^I+\sigma_{\uparrow}^I\sigma_{\downarrow}^I}{\tanh(L/l)(a_IL\sigma_F^I+\sigma_{\uparrow}^I\sigma_{\downarrow}^I)}\\
&=\frac{L}{l}\frac{l\tanh(L/l)\sigma_F^I+\sigma_{\uparrow}^I\sigma_{\downarrow}^IR_IA_T}{\tanh(L/l)(L\sigma_F^I+\sigma_{\uparrow}^I\sigma_{\downarrow}^IR_IA_T)}.
\end{split}
\end{equation}

The second row of Eq.~\eqref{eq:OnsagerRelation} yields the boundary condition for the spin-diffusion equation, whereas the first row in the Onsager relation yields the current through the contact. Multiplying the first row by $\sigma^w W/N_0$, where $W$ is the cross-sectional area of the weak ferromagnet, we obtain
\begin{equation}\label{eq:TunnellingCurrent}
I=G[-\Gamma \gamma V+P_Is_3^w(0)/N_0],
\end{equation}
where 
$$
I=\sigma^w W\partial_z \mu_w(0)
$$
and
$$
G=k_I\sigma^wW.
$$
Since $s_3^w(0)$ is linear in $VN_0$ as shown in Eq.~\eqref{eq:C61Homo}, the spin accumulation contributes an additional resistance to the total resistance.

\bibliography{references}

\begin{thebibliography}{45}
\bibitem [{mat()}]{mathematicanote}%
  \BibitemOpen
  \href@noop {} {}\bibinfo {note} {The Mathematica codes used to solve for the
  coefficients will be provided together with the published paper.}\BibitemShut
  {Stop}%
\end{thebibliography}

\end{document}